\documentclass[preprint,12pt]{elsarticle}

 \usepackage{graphicx}

\usepackage{amsmath,amssymb}
\graphicspath{{./figs/}}

\biboptions{compress}

\journal{Computers \& Fluids}

\begin{document}

\begin{frontmatter}



\title{A Hybrid MPI-OpenMP parallel implementation for pseudospectral
  simulations with application to Taylor--Couette Flow}


\author[a,b]{Liang Shi\corref{cor1}\fnref{fn1}}
\ead{liang.shi@ds.mpg.de}

\author[c]{Markus Rampp}
\ead{markus.rampp@rzg.mpg.de}

\author[a,d]{Bj\"orn Hof}
\ead{bhof@ist.ac.at}

\author[e]{Marc Avila}
\ead{marc.avila@fau.de}

\cortext[cor1]{Corresponding author}
\fntext[fn1]{Present address (visiting scientist): IST Austria, 3400 Klosterneuburg, Austria.}

\address[a]{Max Planck Institute for Dynamics and Self-Organization 
    (MPIDS), 37077 G\"ottingen, Germany}

\address[b]{Institute of Geophysics, University of G\"ottingen, 37077 G\"ottingen, Germany}

\address[c]{Computing Centre (RZG) of the
    Max Planck Society and the Max-Planck-Institute for Plasmaphysics, 
    Boltzmannstr.\ 1, 85748 Garching, Germany}

\address[d]{IST Austria, 3400 Klosterneuburg, Austria}

\address[e]{Institute of Fluid Mechanics, Friedrich-Alexander-Universit\"at Erlangen-N\"urnberg, 91058 Erlangen, Germany}

\begin{abstract}
  A hybrid-parallel direct-numerical-simulation method with application
to turbulent Taylor--Couette flow is presented. The Navier--Stokes
equations are discretized in cylindrical coordinates with the spectral
Fourier--Galerkin method in the axial and azimuthal directions, and
high-order finite differences in the radial direction. Time is
advanced by a second-order, semi-implicit projection scheme, which
requires the solution of five Helmholtz/Poisson equations, avoids
staggered grids and renders very small slip velocities. Nonlinear
terms are evaluated with the pseudospectral method. The code is
parallelized using a hybrid MPI-OpenMP strategy, which, compared with
a flat MPI parallelization, is simpler to implement, allows to reduce
inter-node communications and MPI overhead that become relevant at
high processor-core counts, and helps to contain the memory footprint.
A strong scaling study shows that the hybrid code maintains
scalability up to more than 20\,000 processor cores and thus allows to
perform simulations at higher resolutions than previously feasible. In
particular, it opens up the possibility to simulate turbulent
Taylor-Couette flows at Reynolds numbers up to
$\mathcal{O}(10^5)$. This enables to probe hydrodynamic turbulence in
Keplerian flows in experimentally relevant regimes.
\end{abstract}


\end{frontmatter}


\section{Introduction}\label{sec1}

Rotating fluid flows with radially increasing angular momentum are known to be linearly stable because of the inviscid Rayleigh criterion~\cite{Rayleigh_prsla1917}. A particularly important application is astrophysical Keplerian flow, with  angular velocity profile decreasing radially as $\Omega \propto r^{-3/2}$. Whether Keplerian flows become turbulent because of nonlinear instabilities or remain laminar even at extreme Reynolds numbers, has great implications for accretion processes in weakly-ionized astrophysical disks~\cite{Balbus_nature2011}. This question has been recently investigated with experiments of fluid flows between rotating cylinders (Taylor--Couette flow, TCf), which can in principle approximate Keplerian profiles. Experiments conducted by Ji and co-workers~\cite{JiGoodman_nature2006, SchartmanGoodman_aa2012,edlund2014} found no hydrodynamic turbulence in quasi-Keplerian TCf in the range $Re\sim\mathcal{O}(10^5-10^6)$, whereas similar studies~\cite{PaolettiLathrop_prl2011,PaolettiLathrop_aa2012} report strongly turbulent flows that could account for the observed accretion rates in astrophysical disks. This discrepancy may arise from the axial boundary conditions: numerical simulations of the experimental setups show that top and bottom endwalls confining the fluid strongly disrupt Keplerian velocity profiles and causes turbulence to arise already at $Re\sim\mathcal{O}(10^3)$ \cite{Avila_prl2012}. Hence, the interpretation and extrapolation of experimental data remains controversial because of the prominent role played by axial endwalls. 

Numerical simulations with axially periodic boundaries resolve this problem and allow to directly probe the stability of Keplerian flows. Very recently, Ostilla-M\'onico \emph{et al.} \cite{ostilla2014} have carried out such simulations at $Re=8.1\times 10^4$ and observed that the turbulence decays to laminar flow. However, the role of initial conditions and numerical dissipation in the decay process are unknown and require more detailed studies. Probing the stability of Keplerian flows requires achieving yet larger Reynolds numbers $Re\sim\mathcal{O}(10^5)$, while still resolving the dissipation scales. We note that as the key question is concerned about the existence of turbulence, modeling strategies such as Reynolds-averaged equations (RANS) and Large-Eddy Simulation (LES) are precluded and one has to resort to direct numerical simulation (DNS) of the Navier--Stokes equations (see~\cite{Pope_2000} for details about these simulation techniques). Starting with the study of homogeneous isotropic turbulence conducted by Orszag and Patterson~\cite{OrszagPatterson_prl1972}, DNS has been proven as a very powerful approach to explore the physics of turbulent flows (see Ref.~\cite{MoinMahesh_arfm1998,Jimenez_jot2003}). It has been widely used in fundamental research on both transitional and fully-developed turbulence in boundary layers over a flat plate (e.g. \cite{Spalart_jfm1988,Schlatter_jfm2010}), channel (e.g. \cite{KimMoinMoser_jfm1987,HoyasJimenez_pof2006}), pipe (e.g. \cite{eggels1994,WuAdrian_jfm2012}) and Taylor--Couette flows (e.g. \cite{CoMa96,dong2007,BrauckmannEckhardt_jfm2012}). Distinguished from RANS and LES, a carefully performed DNS resolves all temporal and spatial scales relevant to turbulence and thus provides data of high fidelity. Its advantage is also its main drawback: resolving the physics of turbulence implies a scaling of the computational complexity as $\mathcal{O}(Re^3)$~\cite{Pope_2000}.

In this paper we develop a highly efficient DNS code for TCf with axially periodic boundary conditions using a hybrid two-level parallelization strategy. It enables DNS to be performed up to $Re\sim\mathcal{O}(10^5)$, and thus provides access to a broad range in the parameter space of TCf, including quasi-Keplerian flows at experimentally relevant Reynolds numbers. 
Generally, finite differences \cite{ostilla2014} or spectral-element methods  \cite{dong2007} can be used to perform DNS of Taylor--Couette flow at large Reynolds numbers. However, the most efficient and accurate method for discretizing partial differential equations with periodic boundary conditions is the spectral Fourier--Galerkin method, so we use this in the axial and azimuthal directions. Many authors have also used the spectral Galerkin-method in the non-periodic radial direction by employing Chebyshev, Legendre or Jacobi polynomials ~\cite{Moser_jcp1983,marcus1984,canuto2007}. The two latter render however a computational complexity of $\mathcal{O}(M^2)$, where $M$ is the degree of the approximation, due to the lack of fast transformations between physical and spectral spaces. This makes computations too expensive at large Reynolds numbers. In contrast, with the Chebyshev method the fast cosine transform allows it to keep the cost at $\mathcal{O}(M\log(M))$. However, in order to use accurate quadratures the projection basis must be different from the basis used to discretize the Navier--Stokes equations (Petrov-Galerkin method) ~\cite{Moser_jcp1983,Meseguer_EPJ2007}. On the other hand, if the spectral method is used directly at a collocation grid (in physical space) the resulting differentiation matrices are dense. Hence the solution of the Poisson equations, for example with the diagonalization method \cite{marcus1984}, requires $\mathcal{O}(M^2)$ operations. A common drawback of all the aforementioned spectral methods is that the density of collocation nodes towards the boundaries scales as $\mathcal{O}(M^2)$. Although this allows to properly resolve boundary layers with relatively low resolutions, at large Reynolds numbers the clustering is excessive and the required resolution is often given by the spacing of nodes far from the boundaries. Moreover, this clustering poses a severe restriction on the time step of $\Delta t=\mathcal{O}(M^{-2})$ because of the CFL condition. Although transformations of the node distribution have been proposed~\cite{KasloffTalEzer_jcp1993}, these result in the loss of the spectral convergence. Finally, it becomes impractical to use the Chebyshev method for large resolutions $M\gtrsim 600$, as needed in the simulation of turbulence at large $Re$. For these reasons we use the high-order finite-difference (FD) method in the radial direction, which makes the stretching of grid nodes straightforward. 

The incompressible Navier--Stokes equations in primitive variables are integrated in time with a second-order $\mathcal{O}(\Delta t^2)$ time-splitting scheme proposed by Hugues \& Randriamampianina \cite{Hugues_ijnmf1998}, who tested it in two dimensions in combination with a Chebyshev-Chebyshev discretization. The scheme is semi-implicit and is second-order accurate also for the pressure, rendering a very small $\mathcal{O}(\Delta t^3)$ slip-velocity error at the boundary while  fulfilling the incompressibility constraint. It is straightforward to implement: it avoids staggered grids and requires the solution of five equations of Poisson or Helmholtz type. Raspo \emph{et al.}~\cite{RaspoBontoux_ComptFluids2002}, and later Avila \emph{et al.}~\cite{avila2008}, subsequently extended the scheme to three-dimensional Taylor--Couette flow with no-slip axial boundary conditions, and the Fourier--Galerkin method in the azimuthal direction. These codes have been extensively used for the simulations of centrifugal \cite{czarny2003,czarny2004} and endwall-driven instabilities in Taylor--Couette flow \cite{avila2008,Avila_prl2012}.  Mercader \emph{et al.} \cite{mercader2010} have also extended the scheme to convection in a cylindrical container. Here we combine the scheme of Hugues \& Randriamampianina \cite{Hugues_ijnmf1998} with the finite-difference method in the radial direction and the Fourier--Galerkin method in the azimuthal and axial periodic directions. The nonlinear advective term is computed in physical space with the pseudospectral method. The code is parallelized here by combining the Message Passing Interface (MPI) and the Open Multiprocessing (OpenMP) paradigms. The Fourier-Galerkin method leads to mode-decoupled linear equations, which makes the one-dimensional MPI parallelization rather straightforward to implement. OpenMP threading within MPI tasks allows to efficiently use modern high performance computing (HPC) architectures and mitigates the overhead induced by MPI All-to-all inter-task communications which are typical of spectral methods.

The paper is structured as follows. In \S\ref{sec2}, we formulate the Taylor--Couette problem and then present the numerical method in \S\ref{sec3}. In \S\ref{sec:parallel_implementation} we describe the parallelization strategy employed in the code and its implementation. The accuracy and performance of the code are discussed in \S\ref{sec:validation} and \S\ref{sec:benchmarks}, respectively, before the conclusions in \S\ref{conclusion}.

\section{Governing equations and geometry}\label{sec2}

\begin{figure}[!h]
  \centering
  \includegraphics[width=0.4\textwidth]{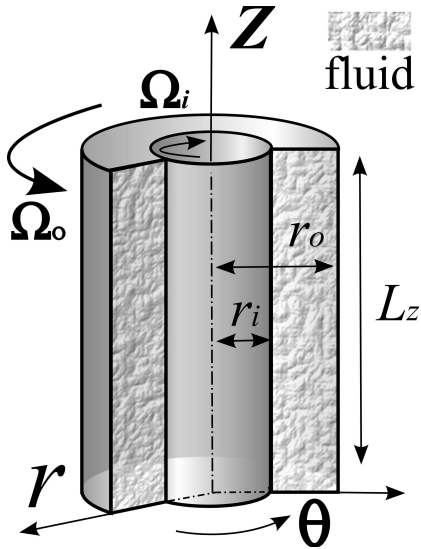}
  \caption{Schematic of the Taylor-Couette system in cylindrical
    coordinates. The inner and outer cylinder rotate independently
    with speeds $\Omega_i$ and $\Omega_o$, respectively.  No-slip
    boundary conditions at the cylinder are used together with axially
    periodic boundary conditions. The fluid between the cylinders
    (hatched region) moves by the shear force due to the fluid
    viscosity.}
  \label{fig:TCgeometry}
\end{figure}

We solve the equations governing the motion of an incompressible fluid
of kinematic viscosity $\nu$ and constant density $\rho$
\begin{equation}
  \partial_{t}\textbf{u}+\textbf{u}\cdot\nabla\textbf{u} =-\frac{1}{\rho}\nabla
  p^h+\nu\Delta\textbf{u},\quad \nabla\cdot\textbf{u}=0,
  \label{eq:NS}
\end{equation}

where $\textbf{u}(\textbf{r},t)$ is the velocity field and
$p^h(\textbf{r},t)$ is the hydrodynamic pressure. Here cylindrical coordinates
$\textbf{r}=(r,\theta,z)$ are used. The geometry of the system is
shown in Fig.~\ref{fig:TCgeometry} and consists of fluid confined
between two concentric cylinders.  The inner (outer) cylinder has
radius $r_{i}$ $(r_o)$ and rotates at a speed of $\Omega_i$
$(\Omega_o)$. The Reynolds number in the inner and outer cylinder is
defined as $Re_{i,o}=\Omega_{i,o}r_{i,o}d/\nu$, where $d=r_o-r_i$ is
the gap between the cylinders. The geometry is fully specified by two
dimensionless parameters: the radii-ratio $\eta=r_i/r_o$ and the
length-to-gap aspect-ratio $\Gamma=L_z/d$, where $L_z$ is the axial
length of the cylinders. At the cylinders no-slip boundary conditions
are applied, whereas in the axial direction periodic boundary
conditions are imposed to avoid endwall effects. This approximates
the case of very long cylinders. In the azimuthal direction periodic
boundary conditions occur naturally. However, it is often
computationally convenient to simulate only an angular section
$L_\theta\le 2\pi$ of the cylinders, and periodic boundary conditions are then
used for $\theta \in [0,L_\theta]$. This is justified provided that the
correlation length of the turbulent flow in the azimuthal direction is
shorter than $r_iL_\theta$ \cite{BrauckmannEckhardt_jfm2012}.

Henceforth, all variables will be rendered dimensionless using $d$,
$\tau=d^2/\nu$, and $\nu^2/d^2$ as units for space, time, and the
reduced pressure $p=p^h/\rho$, respectively. The Navier-Stokes
equations \eqref{eq:NS} for this scaling become
\begin{equation}
  \begin{aligned}
    \partial_{t}\textbf{u}+\textbf{u}\cdot\nabla\textbf{u} &=-\nabla
    p+\Delta\textbf{u}\\
    \nabla\cdot\textbf{u}&=0.
  \end{aligned}
  \label{eq:NS_dimless}
\end{equation}
In cylindrical coordinates the equations read
\begin{equation}
  \begin{aligned}
    (\partial_t+\textbf{u}\cdot\nabla)u_r-u_{\theta}^2/r &= 
    -\partial_rp+\Delta u_r-u_r/r^2-2\partial_{\theta}u_{\theta}/r^2\\
    (\partial_t+\textbf{u}\cdot\nabla)u_{\theta}+u_{\theta}u_r/r &= 
    -\partial_{\theta}p/r+\Delta u_{\theta}-u_{\theta}/r^2
    +2\partial_{\theta}u_r/r^2\\
    (\partial_t+\textbf{u}\cdot\nabla)u_z &= -\partial_zp+\Delta u_z,\\
    u_r/r+\partial_ru_r+\partial_{\theta}u_{\theta}/r+\partial_zu_z &=0.\\
  \end{aligned}
  \label{eq:NS_cyl}
\end{equation}
with $\nabla = (\partial_r, \partial_{\theta}/r,\partial_z)$ and
$\Delta = \partial_r/r+\partial_{rr}^2+\partial_{\theta\theta}^2/r^2+\partial_{zz}^2$.
Note that the Reynolds numbers enter the system through the boundary
conditions
\begin{equation}
  \begin{aligned}
    u_{\theta}(r_{i,o},\theta,z)&=Re_{i,o},\\ u_{r,z}(r_{i,o},\theta,z)&=0,\\
    \textbf{u}(r,\theta,z)&=\textbf{u}(r,\theta + L_\theta,z),\\
    \textbf{u}(r,\theta,z)&=\textbf{u}(r,\theta,z+\Gamma).
  \end{aligned}
  \label{eq:nsBC}
\end{equation}
By taking the divergence of the first equation and then applying
the incompressibility condition, we obtain a Poisson equation for the
pressure,
\begin{equation}
  \Delta p = -\nabla\cdot\textbf{N}(\textbf{u}),\quad \text{where} \quad  
  \textbf{N}(\textbf{u}) = \textbf{u}\cdot\nabla\textbf{u},
  \label{eq:pp}
\end{equation}
with consistent boundary conditions~\cite{GreshoSani_inmf1987}
\begin{equation}
  \partial_np|_{r=r_{i,o}} = \textbf{n}\cdot [-\partial_t\textbf{u}-\textbf{N}(\textbf{u})+\Delta\textbf{u}].
  \label{eq:ppBC}
\end{equation}
As explained in \S\ref{sec3:numericalMethod}, this equation will be
solved for the pressure prediction.


\section{Numerical Formulation}\label{sec3}

The governing equations~\eqref{eq:NS_cyl} are solved for the primitive
variables ($\textbf{u},p$). We discretize the equations with a
combination of the Fourier--Galerkin method with the finite-difference
method (FD) in space, whereas time is advanced with the semi-implicit
fractional-step method of Hugues and
Randriamampianina~\cite{Hugues_ijnmf1998}, who employ
second-order-accurate backward differences with linear (second-order)
extrapolation for the nonlinear term. The pseudospectral technique
with 3/2-dealiasing is applied to
compute the nonlinear term $\textbf{N}(\textbf{u})$ in physical space~\cite{OrszagPatterson_prl1972}.

\subsection{Spatial discretization}

In the periodic axial and azimuthal directions, the velocity field and
pressure are approximated as
\begin{equation}
  \begin{aligned}
    \textbf{u}(r,\theta,z)=\sum_{l=-L}^{L}\sum_{n=-N}^{N}\hat{\textbf{u}}^{ln}(r)
    e^{i(lk_z z+nk_{\theta}\theta)},\\ 
    p(r,\theta,z)=\sum_{l=-L}^{L}\sum_{n=-N}^{N}\hat{p}^{ln}(r)
    e^{i(lk_zz+nk_{\theta}\theta)}, 
  \end{aligned}
  \label{eq:fourier}
\end{equation}
where $k_z$ is the minimum (fundamental) axial wavenumber and fixes
the axial non-dimensional length $\Gamma=2\pi/k_z$ of the
computational domain. Similarly, $L_\theta=2\pi/k_\theta$ is the
azimuthal arc degree; $k_{\theta}=1$ corresponds to the natural periodic
boundary condition in the azimuthal direction, whereas $k_\theta=4$
corresponds to one quarter of an annulus. The hat symbol $\hat{}$ in
\eqref{eq:fourier} denotes quantities in Fourier space and the tuple
$(L,N)$ determines the spectral numerical resolution.

By substituting~\eqref{eq:fourier} into~\eqref{eq:NS_cyl} and
projecting the result onto a basis $e^{-i(lk_z z+nk_{\theta} \theta)}$ 
($l = -L,\dots,L; n=-N,\dots,N$), we obtain the
mode-decoupled Navier-Stokes equations. For each Fourier mode $(l,n)$,
they read
\begin{equation}
  \begin{aligned}
    \partial_t\hat{u}_r+\hat{N}_r &= -\partial_r\hat{p}+\hat{\Delta}\hat{u}_r-
    \hat{u}_r/r^2-2ink_{\theta}\hat{u}_{\theta}/r^2,\\
    \partial_t\hat{u}_{\theta}+\hat{N}_{\theta} &= -ink_{\theta}\hat{p}/r+
    \hat{\Delta}\hat{u}_{\theta}-\hat{u}_{\theta}/r^2-2ink_{\theta}\hat{u}_r/r^2,\\
    \partial_t\hat{u}_z+\hat{N}_z &= -ilk_z\hat{p}+\hat{\Delta}\hat{u}_z.\\
  \end{aligned}
  \label{eq:modalNS}
\end{equation}
Here
$\hat{\Delta}=\partial_r/r+\partial_{rr}-n^2k_{\theta}^2/r^2-l^2k_z^2$,
and the superscripts $(l,n)$ have been omitted for clarity. Note that
the nonlinear term couples Fourier modes and it is thus computed in
physical space with the pseudospectral method. Details of the
implementation and parallelization of the nonlinear term are given in
\S\ref{sec:parallel_implementation}. Equations \eqref{eq:modalNS}
couple the radial and azimuthal velocities. By applying the following
change of variables \cite{orszag1983}
\begin{equation*}
  \begin{aligned}
    \hat{u}_+&=\hat{u}_r+i\hat{u}_{\theta},\\
    \hat{u}_-&=\hat{u}_r-i\hat{u}_{\theta},
  \end{aligned}
\end{equation*}
to equation~\eqref{eq:modalNS}, we obtain the decoupled equations
\begin{equation}
  \begin{aligned}
    \partial_t\hat{u}_+(r)+\hat{N}_+(r) &= -\partial_r\hat{p}(r)+nk_{\theta}\hat{p}(r)/r
    +(\hat{\Delta}-1/r^2-2nk_{\theta}/r^2)\hat{u}_+,\\
    \partial_t\hat{u}_-(r)+\hat{N}_-(r) &= -\partial_r\hat{p}(r)-nk_{\theta}\hat{p}(r)/r
    +(\hat{\Delta}-1/r^2+2nk_{\theta}/r^2)\hat{u}_-,\\
    \partial_t\hat{u}_z(r)+\hat{N}_z(r) &= -ilk_z\hat{p}(r)+\hat{\Delta}\hat{u}_z,
  \end{aligned}
  \label{eq:modalNSdecoup}
\end{equation}
where $\hat{N}_\pm=\hat{N}_r\pm i\hat{N}_{\theta}$.  

We use a standard high-order, central finite-difference method to
approximate the radial derivatives in
equations~\eqref{eq:modalNSdecoup} (see
Ref.~\cite{Fornberg_cambridge}). The radial nodes are distributed as
~\cite{KasloffTalEzer_jcp1993}
\begin{equation}
  r_j= \dfrac{1+\eta}{2(1-\eta)}+\dfrac{\sin^{-1}(-\alpha \cos(\pi
    j/M))}{2\, \sin^{-1}\alpha},\qquad j=0,\dots,M.
  \label{eq:nodeDist}
\end{equation}
For $\alpha=1$ the grid is uniform, whereas for $\alpha\rightarrow 0$
the Chebyshev collocation points are obtained. Here stencils of
$n_s=9$ points, corresponding to a scheme of formally order 8 was
found to give the best compromise in our tests. Note that we reduce
the stencil length gradually towards the boundaries in order to keep
the FD-matrices banded. We show in \S\ref{sec:validation} that due to
the clustering of nodes near the walls with typical values of
$\alpha=0.5$ this reduction of the order of accuracy does not produce
a larger error at the boundaries.

With $(L,N)$ Fourier modes and $M$ radial nodes, the number of grid
points in physical space is $(n_r,n_{\theta},n_z)=(M,2N+1,2L+1)$ in
the radial, azimuthal and axial directions, respectively. Note that we
dealiase the nonlinear term by computing it on a grid of
$(M,3N+1,3L+1)$ points.


\subsection{Temporal scheme}\label{sec3:numericalMethod}

A stiffly stable temporal scheme based on a backward differentiation
formula with extrapolation for the nonlinear term is adopted (see
Ref.~\cite{Hugues_ijnmf1998,KarniadakisOrszag_jcp1991}). It reads
\begin{equation}
  \frac{3\textbf{u}^{i+1}-4\textbf{u}^{i}+\textbf{u}^{i-1}}{2\Delta t} + 
  2\textbf{N}^i(\textbf{u})-\textbf{N}^{i-1}(\textbf{u}) =
  -\nabla p^{i+1} + \Delta{\textbf{u}^{i+1}}.
  \label{eq:timeScheme}
\end{equation}
In the literature this is often referred to as Adams-Bashforth
backward-difference method of second order (AB2BD2). The viscous terms
are discretized implicitly, whereas the nonlinear terms are treated
explicitly. At each time step, equation~\eqref{eq:timeScheme} is
solved through a fractional step method proposed by Hugues and
Randriamampianina~\cite{Hugues_ijnmf1998}. The method is summarized
below. Here $(\hat{\textbf{u}}^i,\hat{p}^i)$ denote the spectral
coefficients at the $i^{th}$ time step.

\begin{enumerate}[\itshape 1\upshape)]
\item 
  Obtain spectral coefficients of the nonlinear term,
  $\hat{\textbf{N}}^{i}(\textbf{u})$, using the 3/2-dealiasing rule
  \begin{itemize}
  \item
    Do matrix-vector multiplication to calculate $\partial_r\hat{\textbf{u}}^i$ (FD method)
  \item Compute dot product in Fourier space to calculate $\partial_{\theta}\hat{\textbf{u}}^i$ 
    and $\partial_z\hat{\textbf{u}}^i$
  \item
    Perform Fourier transform of $\partial_{r,\theta,z}\hat{\textbf{u}}^i$ and $\hat{\textbf{u}}^i$
    to obtain the velocity field and all its derivatives in physical space;
  \item
    Calculate $\textbf{N}^{i}(\textbf{u})=\textbf{u}^i\cdot\nabla\textbf{u}^i$;
  \item
    Perform inverse Fourier transform to obtain the spectral coefficients $\hat{\textbf{N}}^{i}(\textbf{u})$.
  \end{itemize}
\item
  Obtain the pressure prediction, $\hat{p}^*$: solve the Poisson equation
  \begin{align}
    \Delta \hat{p}^* = \nabla\cdot[-2\hat{\textbf{N}}^i(\textbf{u})+\hat{\textbf{N}}^{i-1}(\textbf{u})],
    \label{eq:pressure}
  \end{align}
  with consistent Neumann boundary conditions~\eqref{eq:ppBC}.

\item
  Obtain the velocity prediction, $\hat{\textbf{u}}^*$: solve the three Helmholtz equations 
  \begin{align}
    \frac{3\hat{\textbf{u}}^*-4\hat{\textbf{u}}^{i}+\hat{\textbf{u}}^{i-1}}{2\Delta t} + 
    2\hat{\textbf{N}}^i(\textbf{u})-\hat{\textbf{N}}^{i-1}(\textbf{u}) =
    -\nabla \hat{p}^* + \Delta\hat{\textbf{u}}^*
    \label{eq:vel}
  \end{align}
  with Dirichlet boundary conditions~\eqref{eq:nsBC}. 
\item
  Correct via an intermediate variable $\phi= 2\Delta
  t(\hat{p}^{i+1}-\hat{p}^*)/3$. The incompressibility condition
  $\nabla\cdot \hat{\textbf{u}}^{i+1}=0$ leads to a Poisson equation
  for $\phi$ with homogeneous Neumann boundary conditions (see
  Ref.~\cite{GreshoSani_inmf1987,Hugues_ijnmf1998})
  \begin{equation}
    \begin{aligned}
      &\Delta \phi = \nabla\cdot\hat{\textbf{u}}^*,\\
      &\partial_r\phi|_{r=r_{i,o}} =0
      \label{eq:correction}
    \end{aligned}
  \end{equation}
\item
  Compute pressure and velocity correction, $\hat{p}^{i+1}$ and $\hat{\textbf{u}}^{i+1}$:
  \begin{equation}
    \begin{aligned}
      \hat{p}^{i+1}&=\hat{p}^*+3\phi/(2\Delta t)\\
      \hat{\textbf{u}}^{i+1}&=\hat{\textbf{u}}^*-\nabla\phi    
    \end{aligned}
    \label{eq:update}
  \end{equation}
\item
  Go back to step 1
\end{enumerate} 

The Navier-Stokes equations are thus advanced in time by solving five
systems of linear equations \eqref{eq:pressure}-\eqref{eq:correction},
of Poisson or Helmholtz type, for each Fourier mode. This method
accounts for a divergence-free velocity field and a small slip at the
wall of the order of $\mathcal{O}(\Delta t^3)$ in the tangential
velocities, $u_z$ and $u_\theta$. We note that the method was
originally developed and tested \cite{Hugues_ijnmf1998} for the
two-dimensional Navier-Stokes equation discretized on a
Chebyshev-Chebyshev collocation grid, and the Poisson and Helmholtz
equations were solved using the double diagonalization method, thus
rendering quadratic computational complexity in each direction. Here,
the FD-discretized Poisson and Helmholtz equations render banded
matrices which are solved with the LU-method. The decompositions are
precomputed at the beginning of a simulation and at each time step
only backward and forward substitutions need to be computed, resulting
in an operation count of $\mathcal{O}(M)$ for the solution of each
system. Note that for the axially and azimuthally invariant Fourier
mode, $n=l=0$, the Poisson equations \eqref{eq:pressure} and
\eqref{eq:correction} are singular: their solution is defined up to a
constant because of the Neumann boundary conditions. Here a Dirichlet
homogeneous boundary condition was employed at the outer cylinder to
select a particular solution.


\section{Parallelization scheme and its implementation}\label{sec:parallel_implementation}

\begin{figure}[!ht]
  \centering
  \includegraphics[width=1.0\textwidth]{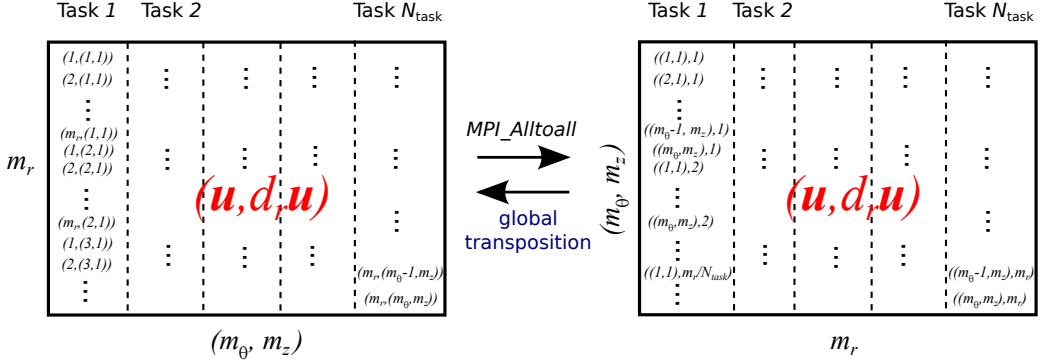}
  \caption{Schematic of the MPI-based, one-dimensional ``slab'' domain decomposition and the global 
    transposition by using the function {\tt MPI\_Alltoall()}.
    Mode-independent spectral coefficients in Fourier space are 
    distributed among different MPI tasks. Each variable has a dimension of $(m_r,m_{\theta},m_z)$.}
  \label{fig:mpiDist}
\end{figure}

A hybrid MPI-OpenMP parallelization strategy is adopted for the implementation of the 
code. Since the linear equations \eqref{eq:pressure}--\eqref{eq:correction} are mode-independent, it is 
convenient to employ an MPI-based, one-dimensional domain decomposition (also known as ``slab'' 
decomposition, Fig.~\ref{fig:mpiDist}): 
The Fourier coefficients $(\hat{u}_+,\hat{u}_-,\hat{u}_z,\hat{p})$ corresponding to
different modes are distributed across the MPI tasks, which allows to solve 
equations \eqref{eq:pressure}--\eqref{eq:correction} concurrently, without inter-task communications.
Each of the $N_\mathrm{tasks}$ MPI tasks operates on data corresponding to a number of 
$m_{\theta}\cdot m_z/N_\mathrm{tasks}$ modes, where $(m_r,m_{\theta},m_z)=(M,N+1,2L)$ are the 
dimensions of variables in Fourier space.
OpenMP threading inside each MPI task allows to efficiently exploit
the remaining coarse-grained parallelism (see below).

We compute the nonlinear term (step \emph{1} in Section~\ref{sec3:numericalMethod}) by performing
global matrix transpositions (Fig.~\ref{fig:mpiDist}) of the discretized 
fields $\partial_r\hat{\textbf{u}}$ and $\hat{\textbf{u}}$ such that
for each radial point the complete spectrum of Fourier modes is localized in one MPI task. 
This requires a collective communication operation of type
"all-to-all" but allows to most efficiently compute the Fourier transformations and 
the derivatives with respect to the spectral coordinates, namely $\theta$ and $z$. 
Finally, inverse transpositions are performed for the resulting
array $\hat{\textbf{N}}$. 

In our applications, typically $m_{\theta}\cdot m_z \gg m_r$ applies,
i.e.\ there are many more Fourier modes than radial grid points.
Hence, the number of MPI tasks in our slab decomposition
is bounded by $N_\mathrm{tasks} \le m_r$ (cf.~Fig..~\ref{fig:mpiDist})
and consequently the achievable parallel
speedup with respect to the serial code would be at most $m_r$. However, OpenMP threads allow to parallelize over the 
$m_{\theta} \cdot m_z/N_\mathrm{tasks}$ modes within a MPI task, while retaining the
one-dimensional MPI domain decomposition, which is conceptually 
straightforward to implement. Similarly, we can exploit
concurrency in the nonlinear part if $N_\mathrm{tasks} < m_r$ applies. 
In addition, the Fourier transformations and the individual partial derivatives required for evaluating
$\textbf{u}\cdot\nabla\textbf{u}$ are computed concurrently and the transposition of 
$\partial_r\hat{\textbf{u}}$ is overlapped with the computation of 
$\textbf{u}$, $\partial_{\theta}\textbf{u}$, and $\partial_{z}\textbf{u}$.

Theoretically, this strategy allows to utilize a number of $\min(m_r,m_{\theta}\cdot m_z)\cdot
N_\mathrm{threads}$ processor cores where $N_\mathrm{threads}$ is the maximum number of threads a shared-memory compute node
provides. Current high performance computing (HPC) platforms feature
at least 16 cores with 32 logical threads per node (e.g. Intel Xeon E5
Sandy-Bridge or Ivy-Bridge processors), and thread-based concurrency on the node-level is expected to
increase substantially in the near future, in particular with the
many-core processors and GPU-accelerated nodes  
\cite{dongarra_hpc2011}. 
In practice, we achieve the best parallel efficiencies when MPI tasks are mapped to
the individual "sockets" (i.e.\ CPUs or NUMA domains) of a compute node and the
number of OpenMP threads equals the number of physical cores per
socket. Due to the smaller number of MPI tasks per node (compared with a plain
MPI parallelization) the amount of inter-node communications is reduced in the global
transposition. This transposition, which is implemented by {\tt MPI\_Alltoall} collective
communication and task-local transpositions, ultimately limits the
overall parallel scalability of the code at high task counts (see\ Section~\ref{sec:benchmarks}).

The code is implemented in FORTRAN~90 and has been ported to a number of major HPC architectures,
including IBM Power and BlueGene, as well as compute clusters based on 
x86\_64 processors and high-performance interconnects such as InfiniBand.
We employ vendor-optimized BLAS and LAPACK routines for the 
matrix-vector multiplication (BLAS level-2 routine {\tt DGEMV}) and the linear solvers 
(LAPACK routines {\tt DGBTRF}, {\tt DGBTRS} taken e.g.\ from
the Intel Math Kernel Library, MKL, or IBM ESSL), respectively, and
utilize the MKL or the FFTW library \cite{FFTW} for performing the Fourier
transformations in the nonlinear part of the code.  
For data output we employ the parallel HDF5
  libraries which enable collective output of the MPI-distributed data into a
  single file in a transparent and efficient way. This facilitates 
  data handling, post-processing and visualization, \textit{e.g.} with VisIT 
  or Paraview (cf.\ Fig.~\ref{fig:highReRun}).


\section{Numerical Accuracy and Code Validation}\label{sec:validation}

The code has been tested\footnotemark[1] over a wide range of Reynolds numbers $Re\in[50,100\,000]$. 
A number of specific test cases will be given in the following. \footnotetext[1]{The platform for the tests is a small departmental cluster with two Intel Xeon E5640 four-core processors 
in each node. The Intel Compiler (v12.0) and the Intel MKL library
(v10.3) were employed. The experiments described in the last part in this section (fully turbulent flow) were done on the same platform as described in \S6.}

\subsection{Laminar flow}

We firstly computed the laminar velocity profile, which is also known as circular Couette flow. It can be expressed as $\textbf{U}=(0,U_{\theta}(r),0)$, where $U_{\theta}(r)=C_1r+C_2/r$ with $ C_1= (Re_o-\eta Re_i)/(1+\eta)$ and $C_2= \eta(Re_i-\eta Re_o)/((1-\eta)(1-\eta^2))$, and corresponds to pure rotary shear flow. The tests were performed at $Re_i=50, Re_o=200$ and at $\eta=0.5$. 
\begin{figure}[!ht]
  \centering
  \includegraphics[width=0.7\textwidth]{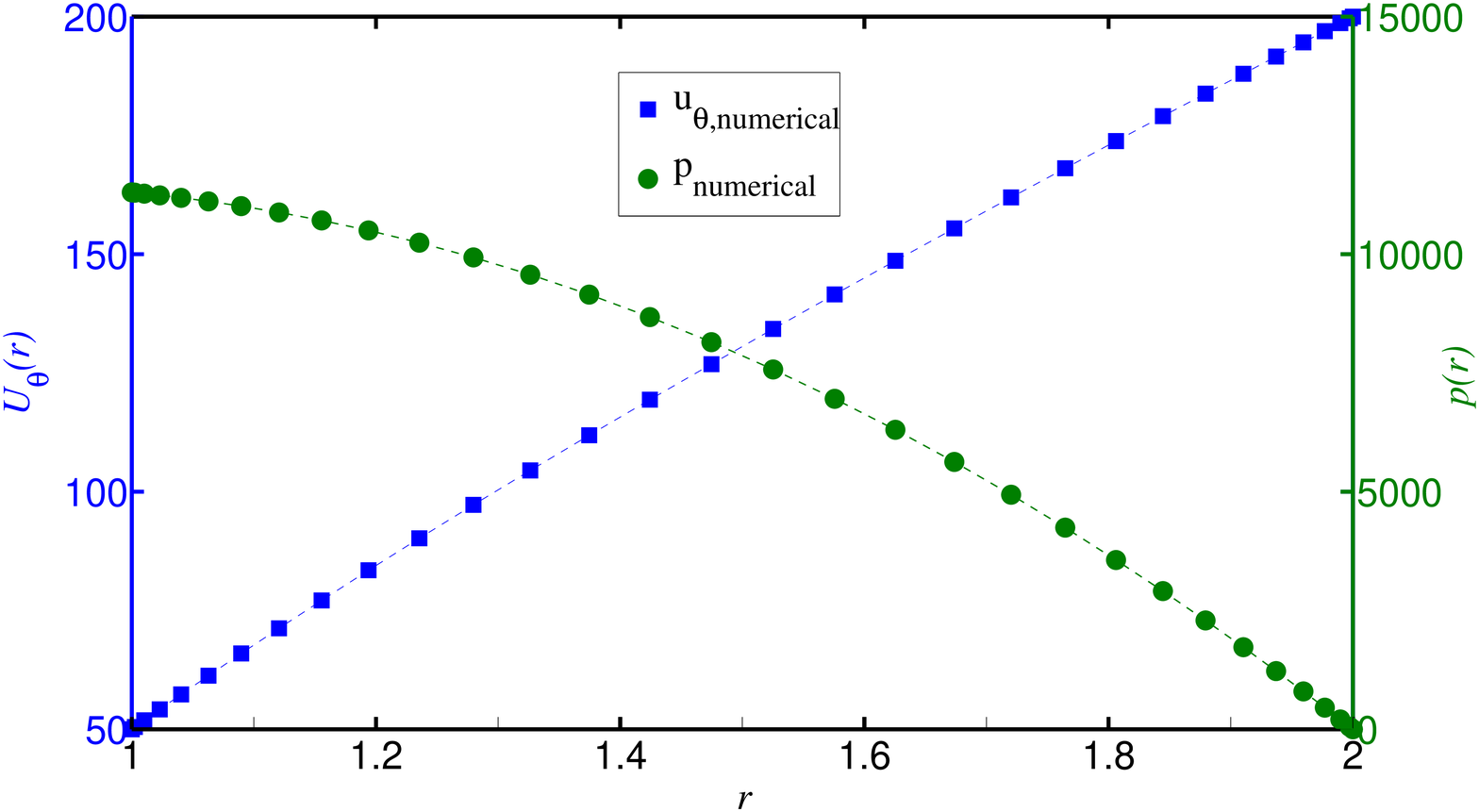} \\
  \includegraphics[width=0.7\textwidth]{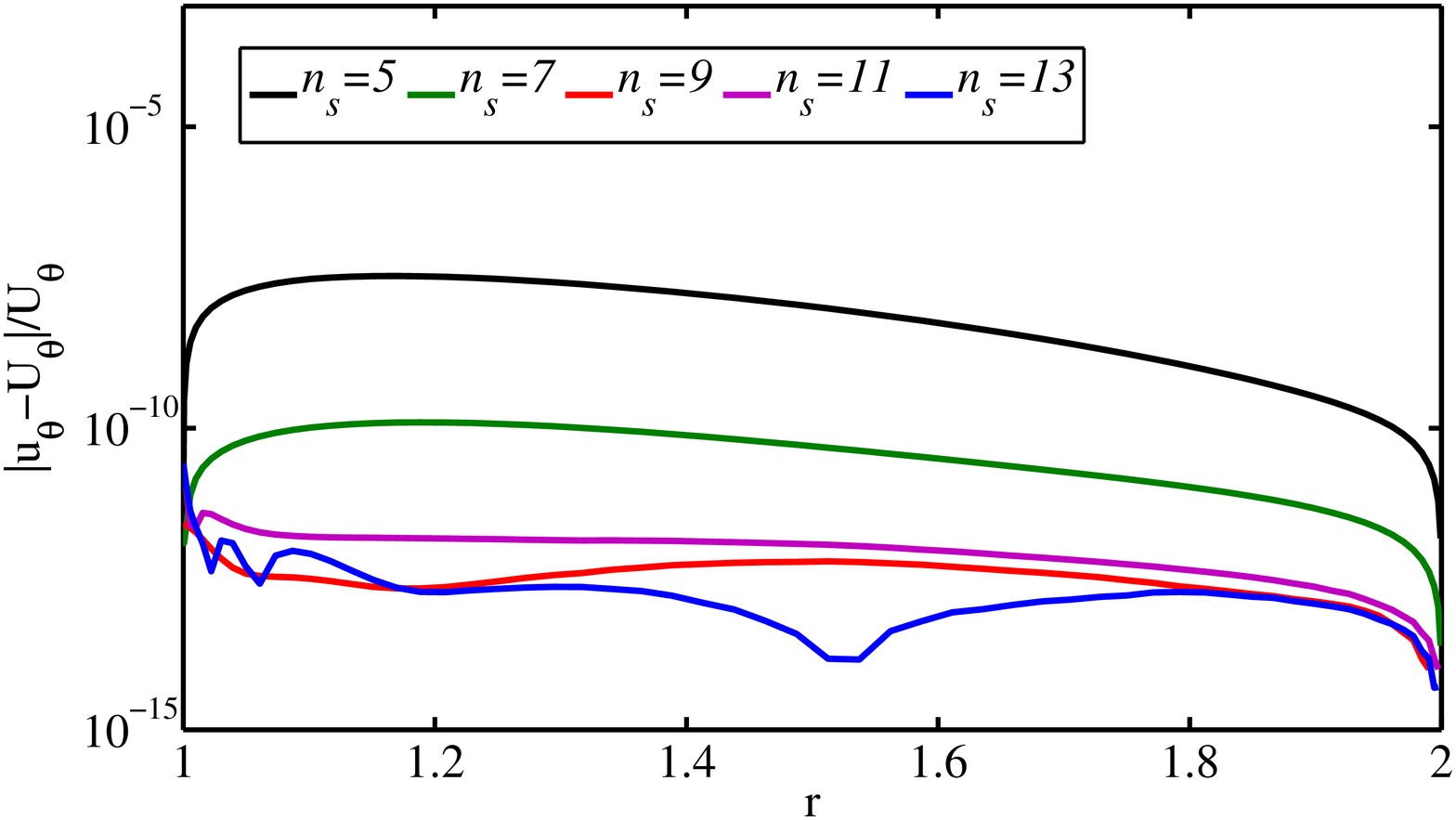}\\
  \caption{Laminar Couette flow at $Re_i=50, Re_o=200$ and $\eta=0.5$
    with Chebyshev points ($\alpha\rightarrow 0$ in equation
    \eqref{eq:nodeDist}). Top: numerically obtained streamwise
    velocity profile (blue squares) and pressure (green circles) for
    $n_r=32$. The dashed lines show the corresponding curves for the
    exact Couette solution. Bottom: local relative error $\epsilon_u$
    for $n_r=64$ as a function of $r$ and several stencil lengths
    $n_s$.}
  \label{fig:lamProfile}
\end{figure}
A non-uniform grid according to formula (\ref{eq:nodeDist}) was used in the radial direction. Fig.~\ref{fig:lamProfile}(top) shows the numerical velocity and pressure profiles for $\alpha\rightarrow 0$ (Chebyshev points) and $n_r=32$, which match well with the theoretical curves (dashed lines). The distributions of the relative error $\epsilon_u(r)=|\frac{u_{\theta}-U_{\theta}}{U_{\theta}}|$ along the radial direction are shown in Fig.~\ref{fig:lamProfile}(bottom) for $n_r=64$ and different stencil lengths $n_s$. In the FD method, the stencil length is the number of consecutive points used to approximate the derivatives. As $n_s$ is increased, the relative error decreases until approaching  the machine precision. We found that a stencil of 9 points gives a good compromise between computing time and accuracy, so $n_s=9$ is kept for the following tests.

\begin{figure}[!ht]
  \centering
  \includegraphics[width=0.49\textwidth]{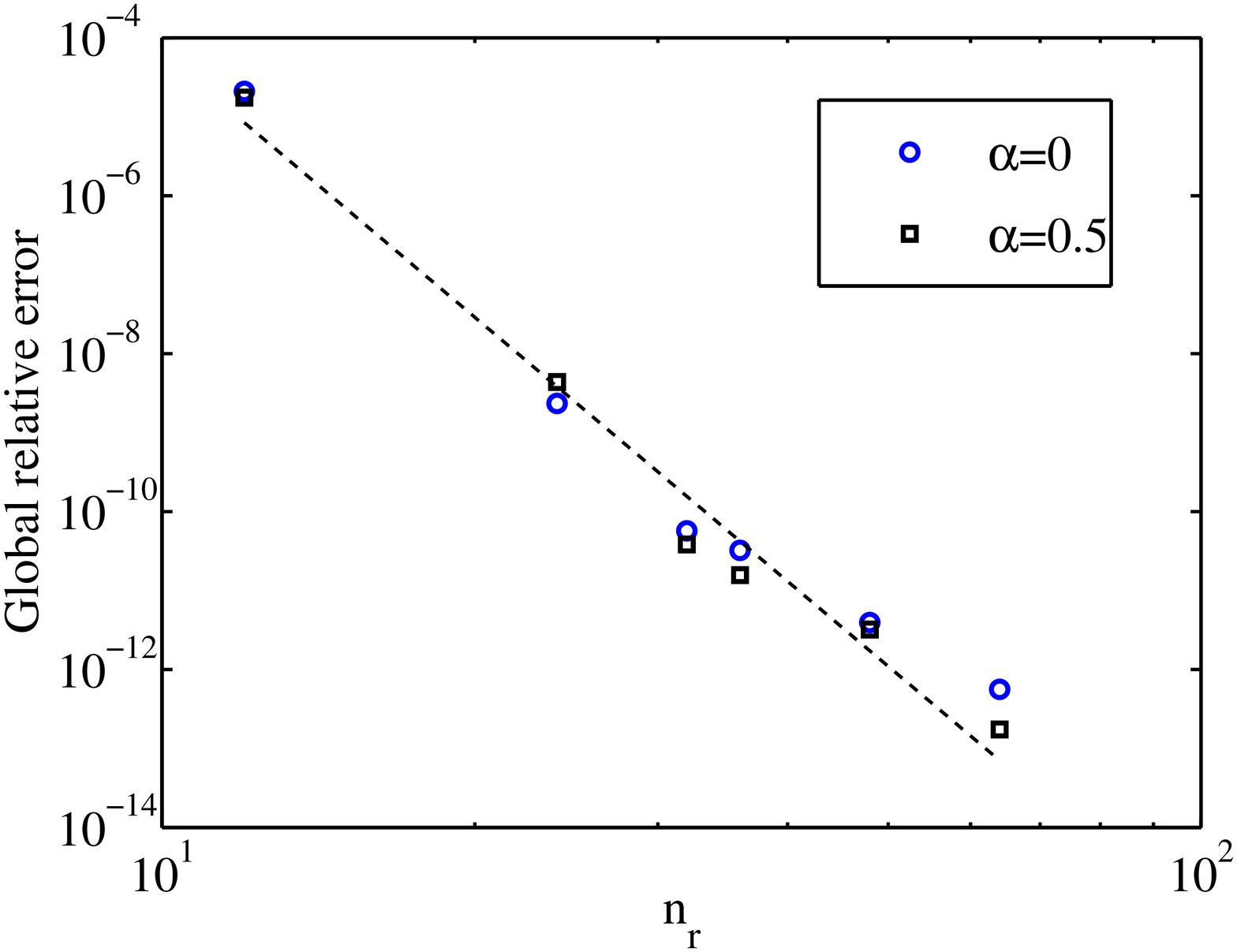}
  \includegraphics[width=0.47\textwidth]{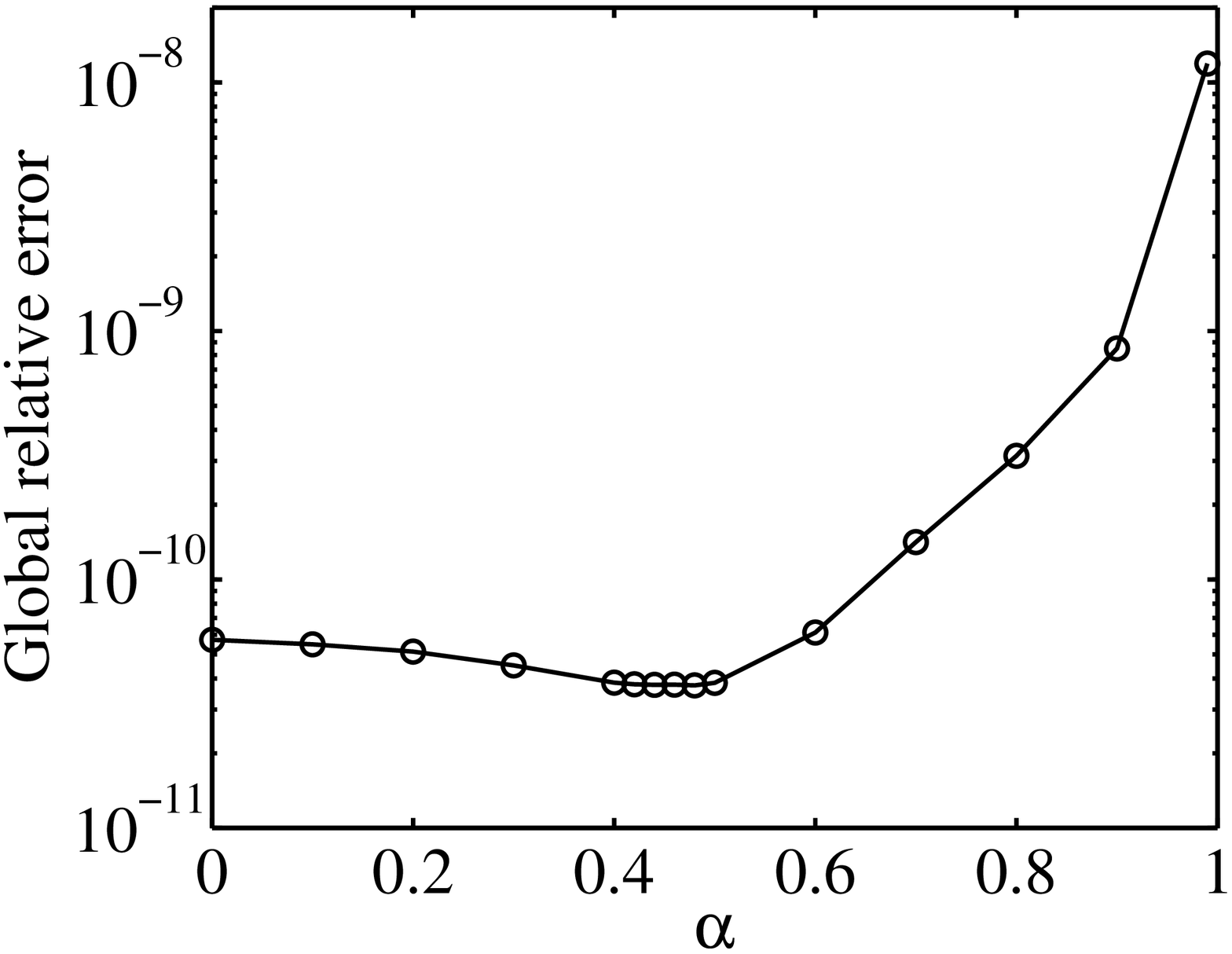}
  \caption{Global relative error $\mathcal{E}_u$ as a function of
    $n_r$ (left) and $\alpha$ for $n_r=32$ (right). The dashed line
    in the left panel is a power fit with an exponent of -11. The
    stencil length is $n_s=9$ in both panels.}
  \label{fig:relErrE}
\end{figure}

To measure the global error, we integrated the local error $\epsilon_u$ over the radial direction, $\mathcal{E}_u=\int_{r_i}^{r_o} \epsilon_u r dr$. This is shown in Fig.~\ref{fig:relErrE} as a function of $n_r$ and $\alpha$.  In the left panel, $\mathcal{E}_u$ scales as a power law with $n_r$ for both $\alpha=0$ and $\alpha =0.5$. The power exponent is about $-11$, which is better than expected from the 9-point-stencil FD scheme. The right panel shows that the error is minimized for $\alpha\simeq 0.5$ and that below 0.5 the errors are almost at the same level.

\subsection{Hydrodynamic instability and three-dimensional time-dependent flow}

As the Reynolds number of the inner cylinder increases beyond a
certain value, laminar Couette flow gives way to Taylor vortices, and
subsequently to wavy vortex flow. Following Jones~\cite{Jones_jfm1985}
we computed the onset of wavy vortex flow for $\eta=0.56$,
$\Gamma=2.2$ and $k_\theta=1$. The simulations were
initialized with Taylor vortex flow, which was disturbed with a
perturbation of azimuthal wavenumber $n$. We then measured the
exponential growth/decay rate of the kinetic energy of the disturbed
mode, which vanishes at the critical point. In agreement with Jones we
found that the dominant mode has $n=1$. The critical Reynolds number, obtained with $(n_r,n_\theta,n_z)=(64,48, 128)$ and $\Delta t = 2\times 10^{-5}$, was determined to
$Re_i^{c}\in [408.09,408.1]$, which was reproduced by using the
Petrov--Galerkin code of Meseguer \emph{et
  al.}~\cite{Meseguer_EPJ2007}. We note that Jones quotes a slightly
lower value (399, Table~1 of~\cite{Jones_jfm1985}). We attribute this
discrepancy of about 2\% to the limited axial and radial resolutions,
which Jones could use in the eighties.

\begin{figure}[!ht]
  \centering
  \includegraphics[height=0.3\textwidth]{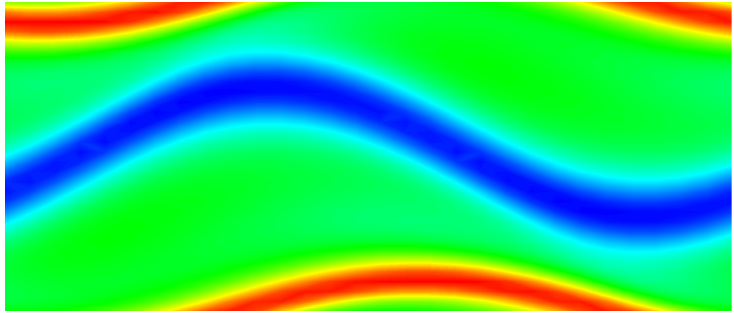}\\[5pt]
  \caption{Contour plots of the streamwise velocity in the middle
    $(\theta,z)$ plane for wavy Taylor vortices.  The outer cylinder
    is stationary, whereas the inner cylinder rotates with
    $Re_i=458.1$. The geometrical parameters are $\eta=0.868$ and
    $\Gamma=2.4$ and only one sixth of the circle ($k_\theta=6$) was
    used in the simulations and is displayed. Here
    $(n_r,n_\theta,n_z)=(32, 32, 32)$.}
  \label{fig:taylorVortices}
\end{figure}

Nonlinear time-dependent wavy vortex flow was computed at $Re_i=458.1,
Re_o=0$, $\eta=0.868$ and is shown in
Fig.~\ref{fig:taylorVortices}. The axial length was chosen as
$\Gamma=2.4$ and $k_{\theta}=6$ to compare to the experimental
observations of King \emph{et al.}~\cite{King_jfm1984} and numerical
simulations of Marcus~\cite{marcus1984}. Wavy Taylor vortices are a
relative equilibrium: they consist of a constant pattern rotating as a
solid at a constant wave speed. Marcus~\cite{marcus1984} notes:
`\emph{A test that is more sensitive than the comparison of torques is
  the comparison of the numerically computed wave speed with the
  experimentally observed wave speed}'. We performed this test with
spatial resolution $(n_r,n_\theta,n_z)=(32, 32, 32)$ and
time-step size $\Delta t=2\times 10^{-5}$. The wave speed normalized
by the rotation speed of the inner cylinder was accurately computed
with a rigorous method based on Brent's minimization
algorithm~\cite{Brent_Dover}. Our result, namely a wave speed of
$c=0.34432$, with the pattern rotating at about one-third of the speed
of the inner cylinder, agrees to all decimal places given
in~\cite{King_jfm1984}. The same result was reproduced with higher resolutions (as in \S\ref{sec:slip}) and on 
various HPC platforms.

The choice $k_\theta=6$, as in \cite{marcus1984}, automatically fixes the symmetry of the observed wavy mode. To remove this constraint, we did an extra set of  simulations with $k_{\theta}=1$, {\em i.e.} the whole annulus in the azimuthal direction. After initial transients the flow stabilized to a wavy state, whose wavenumber and speed depended on the disturbance used to initialize the run. When a disturbance with $n=6$ was used, exactly the same result as above was obtained. We also tried $n=5$ an $n=7$ and obtained new wavy-vortex-flow sates, all of which had very similar wavespeed. The same results were reproduced by doubling the axial  length of the domain, i.e. $\Gamma=4.2$, and for much longer cylinders ($\Gamma=14.4$). Finally, we note that for other disturbances, which simultaneously excited several modes, we could also obtain chaotic flow states. Overall, the results are in agreement with the experiments of Coles \cite{coles1965}, who demonstrated the coexistence of several flow states at identical parameter values depending on the history of the flow. An investigation of the basin of attraction of each of the possible flow states would be very interesting and may be pursued in the future. We expect extreme multiplicity, much beyond what Coles observed.

\subsection{Velocity slip at the cylinders}\label{sec:slip}

\begin{figure}[!ht]
  \centering
  \includegraphics[width=0.7\textwidth]{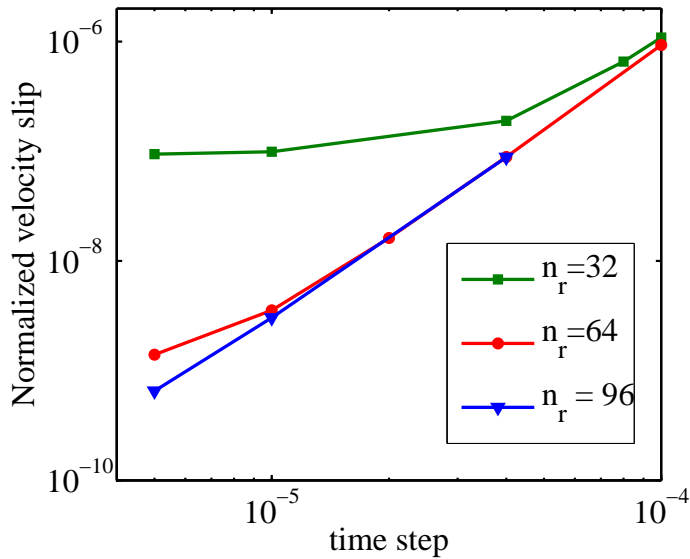}
  \caption{Normalized velocity slip at the inner cylinder versus
    time-step size $\Delta t$ for different $n_r$. The parameters are
    the same as in Figure~\ref{fig:taylorVortices}, corresponding to wavy vortex flow. The spatial resolution in the
    axial and azimuthal directions is $(n_\theta,n_z)=(32,32)$}
  \label{fig:slipVel}
\end{figure}

We further examined the tangential velocity slip at the cylinders. In the projection scheme we employed, the incompressibility constraint $\nabla\cdot \textbf{u}=0$ is discretely fulfilled by construction, in that the Poisson equation for $\phi$ in \S\ref{sec3:numericalMethod} is derived by applying the divergence-free condition. However, the velocities at the inner and outer cylinders slip by a predicted amount of $|\nabla \phi|=\mathcal{O}(\Delta t^3)$ after the correction step~\cite{Hugues_ijnmf1998,RaspoBontoux_ComptFluids2002}. We have evaluated the L2-norm of the tangential velocity slip at the inner cylinder, $\int_{\theta}\int_{z}\sqrt{((u_{\theta}-Re_i)^2-u_z^2)|_{r=r_{i}}}d\theta dz$.
In Fig.~\ref{fig:slipVel} the relative velocity slip, i.e. slip velocity normalized with $Re_i$, is shown as a function of $\Delta t$ for several radial resolutions $n_r$. For the lowest resolution $n_r=32$ the curve rapidly levels off, indicating that spatial-discretization errors dominate over temporal errors. Note that with the largest time-step size allowed for stability and lowest resolution we already obtain five digits in the accuracy of $c$. As $n_r$ is increased the slip velocity decreases and its scaling gradually approaches a power law, here with an exponent of approximately 2.5. The reason why this is slightly below the predicted value of $3$, as observed by Raspo \emph{et al}. \cite{RaspoBontoux_ComptFluids2002} for simple cosine and sine flows, is unclear. The behaviour seen in Fig.~\ref{fig:slipVel} suggests that very high spatial resolutions may be needed to observe the asymptotic scaling. Nevertheless we stress that the even with the coarsest resolution and time step, the relative slip error is of about $10^{-6}$. We repeated this study for fully turbulent flow at $Re_i=8000$ (see \S\ref{sec:turb} for further tests with the same setup). The results were very similar, and in fact, even with $\Delta t =2\times 10^{-7}$, which is close to the stability limit ($\Delta t_\text{lim}\approx 3\times 10^{-7}$), the relative error was $10^{-9}$. We conclude that in typical simulations the dominating source of error comes from the spatial discretization, which determines the slip error observed in practice. A time step moderately smaller than permitted by stability yields very accurate results in the solution and very small slip velocities.

\subsection{Localized turbulence at moderate Re}\label{sec:stripe}

\begin{figure}[!ht]
  \centering
  \includegraphics[width=0.75\textwidth]{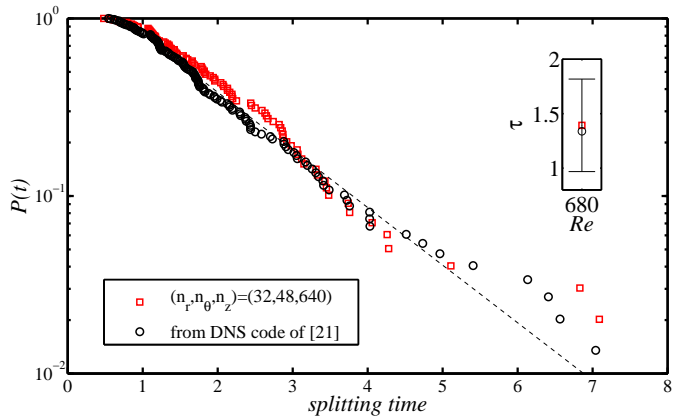}
  \caption{Probability distributions of the splitting time of a single
    turbulent stripe at $Re_i=680,Re_o=-680$ and at
    $\eta=0.993$. Circles correspond to the data set obtained with the
    spectral Petrov--Galerkin code of Meseguer \emph{et
      al.}~\cite{Meseguer_EPJ2007}, whereas squares correspond tot the
    data set obtained with the present code with
    $(n_r,n_\theta,n_z)=(32,48,640)$. Inset: characteristic splitting
    time estimated with the sample mean. The error bar shows the
    $95\%$ confidential interval.}
  \label{fig:resolComp}
\end{figure}

Localized turbulence, interspersed in the surrounding laminar flow, is
a typical feature of transitional Reynolds numbers in shear flows. The
turbulent stripe pattern found in the counter-rotating Taylor-Couette
flow in the narrow-gap limit is an example. We computed this pattern
for exact counter-rotation ($Re_i=-Re_o=680$) and $\eta=0.993$. The
time-step size was $\Delta t = 2\times 10^{-5}$ and the domain size in
the axial direction was $\Gamma=50$, whereas $k_{\theta}=
179$. Folowing Barkley and Tuckermann~\cite{barkley2005} the
$\theta$-direction in our computational domain is tilted with an angle
of $24^{\circ}$ to the streamwise direction (for details
see~\cite{ShiHof_prl2013}). We tested the probability distributions of
the splitting times of turbulent stripes reported
in~\cite{ShiHof_prl2013}, which were obtained by using the spectral
Petrov-Galerkin code of Meseguer \emph{et
  al.}~\cite{Meseguer_EPJ2007}. We here used $n_r=32$ in the radial
direction, whereas Shi \emph{et al.}~\cite{ShiHof_prl2013} used
modified Chebyshev polynomials of degree up to 26. In both cases the
azimuthal and axial resolutions are $n_\theta=48$ and $n_z=640$,
respectively. The exponential distributions of splitting times
obtained by both codes are statistically equivalent (see the inset in
Fig.~\ref{fig:resolComp}): our computed characteristic time of the
exponential distributions is well within the $95\%$ confidence
interval reported in~\cite{ShiHof_prl2013}.

\subsection{Fully turbulent flow at high Re}\label{sec:turb}

\begin{figure}[!ht]
  \centering
  \includegraphics[width=0.75\textwidth]{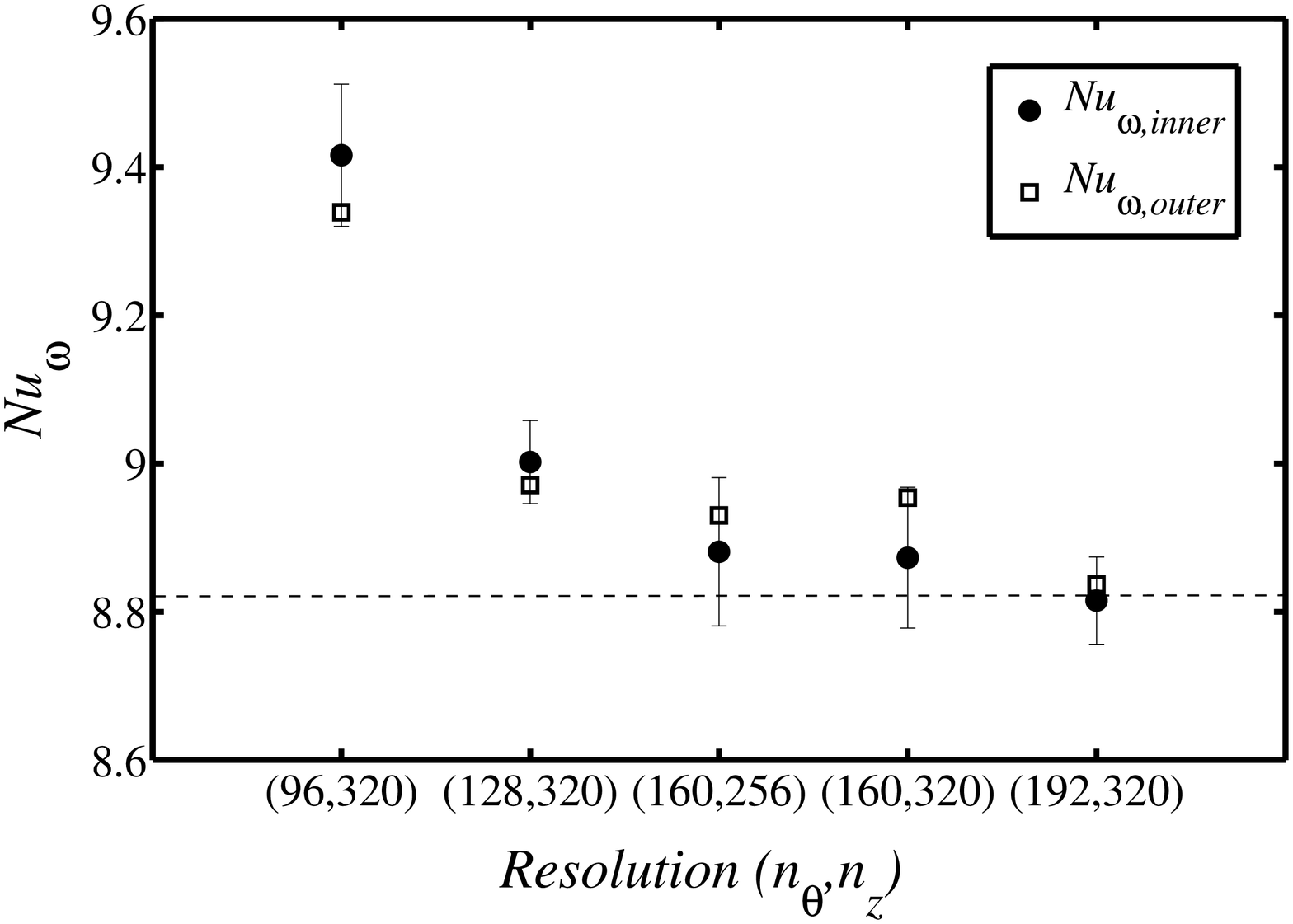}\\
  \includegraphics[width=0.75\textwidth]{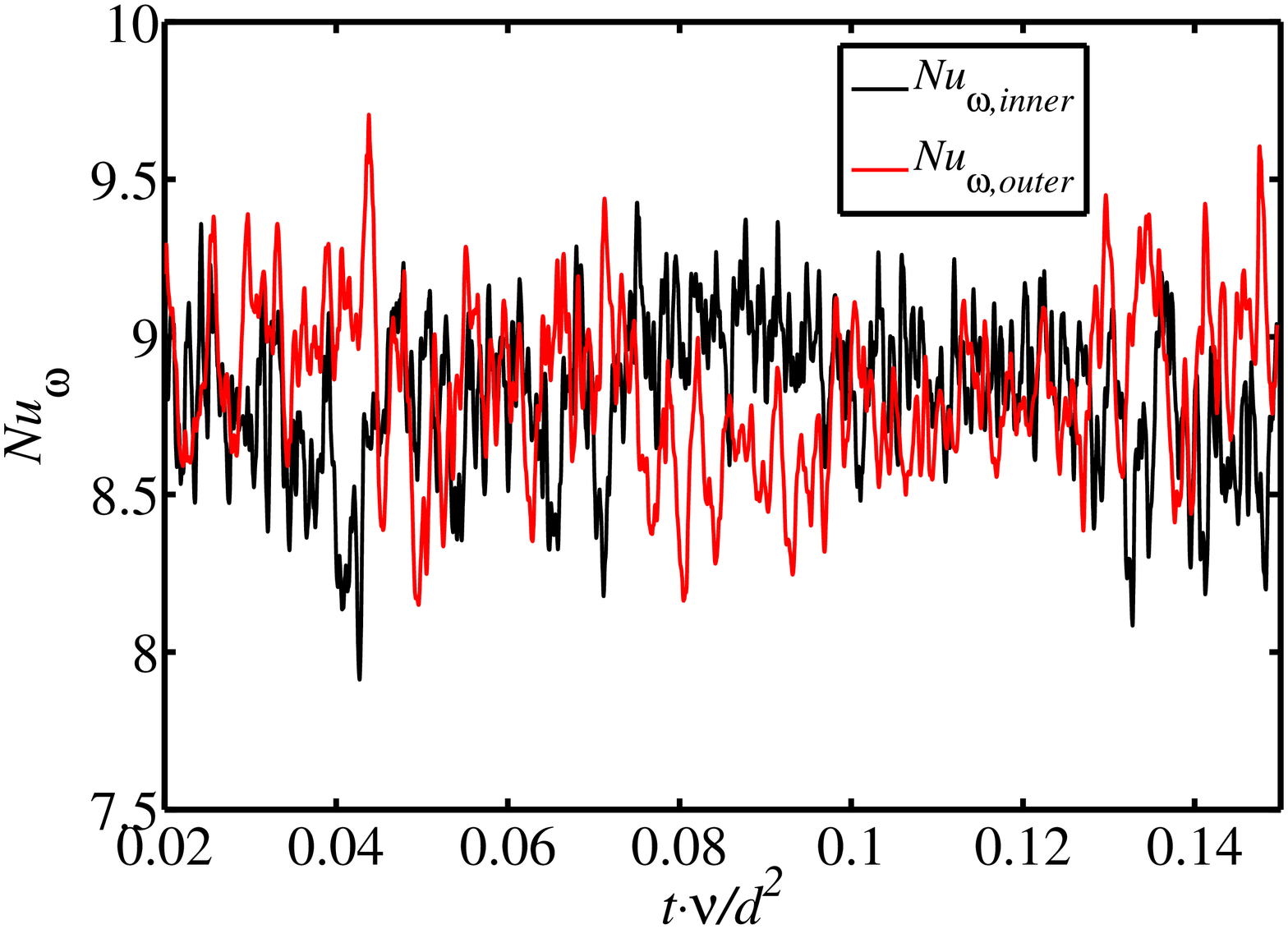}
  \caption{(Top) Quasi-Nusselt number at $Re_i=8000$, $Re_o=0$,
    $\eta=0.50$, $\Gamma=2\pi/k_z=\pi$ and $k_\theta=2$, as a function of the
    azimuthal and axial resolutions for $n_r=128$. The dashed line
    corresponds to the value ($Nu_{\omega}\simeq 8.816$) reported
    in~\cite{BrauckmannEckhardt_jfm2012}. The error bars indicate the
    $95\%$ confidential interval. (Bottom) The temporal fluctuation of
    the quasi-Nusselt number at the highest resolution $(n_r,n_\theta,n_z)=
    (128,192,320)$.}
  \label{fig:nuVSresol}
\end{figure}

The robustness of the code was further validated at high Reynolds
numbers in the linearly unstable regime, where the flow is fully
turbulent. Here we computed the global torque exerted by the fluid on
the inner and outer cylinders, which characterizes the turbulence
intensity and the transport of angular
momentum~\cite{BrauckmannEckhardt_jfm2012}. The tests were done at
$Re_i=8000$ and stationary outer cylinder with $\eta=0.5$, $\Gamma =
2\pi/k_z=\pi$ and $k_{\theta}=2$. The time-step size is $\Delta t =
2\times 10^{-7}$. As is shown in Fig.~\ref{fig:nuVSresol}(top), the
quasi-Nusselt number $Nu_{\omega}$~\cite{BrauckmannEckhardt_jfm2012},
which is the torque normalized by the torque of the laminar flow,
converges to $8.815$ at the resolution
$(n_r,n_{\theta},n_z)=(128,192,320)$. This value agrees very well to
the value of $8.816$ recently reported by Brauckmann and
Eckhardt~\cite{BrauckmannEckhardt_jfm2012}, who also used the spectral
Petrov-Galerkin code of Meseguer \emph{et
  al.}~\cite{Meseguer_EPJ2007}.  The temporal fluctuation of the
quasi-Nusselt number obtained with the highest resolution is shown in
the bottom figure. At this $Re$, we also examined the influence of the
radial node distribution by varying the parameter $\alpha$ in
equation~\eqref{eq:nodeDist}. Three runs with $\alpha=0,0.5,0.99$
were done, and all three rendered $Nu_{\omega}=8.81\pm 0.05$.  The
maximum time-step size for stability was found to be  $\Delta t_\text{lim} \approx 3\times 10^{-7}$ for all three values of $\alpha$. This is explained by the fact that
the CFL number is dominated by the azimuthal direction. Thus at this
Reynolds number the Chebyshev node distribution does not impose a
restriction in the time-step yet.

\begin{figure}[!ht]
  \centering
  \includegraphics[width=0.9\textwidth]{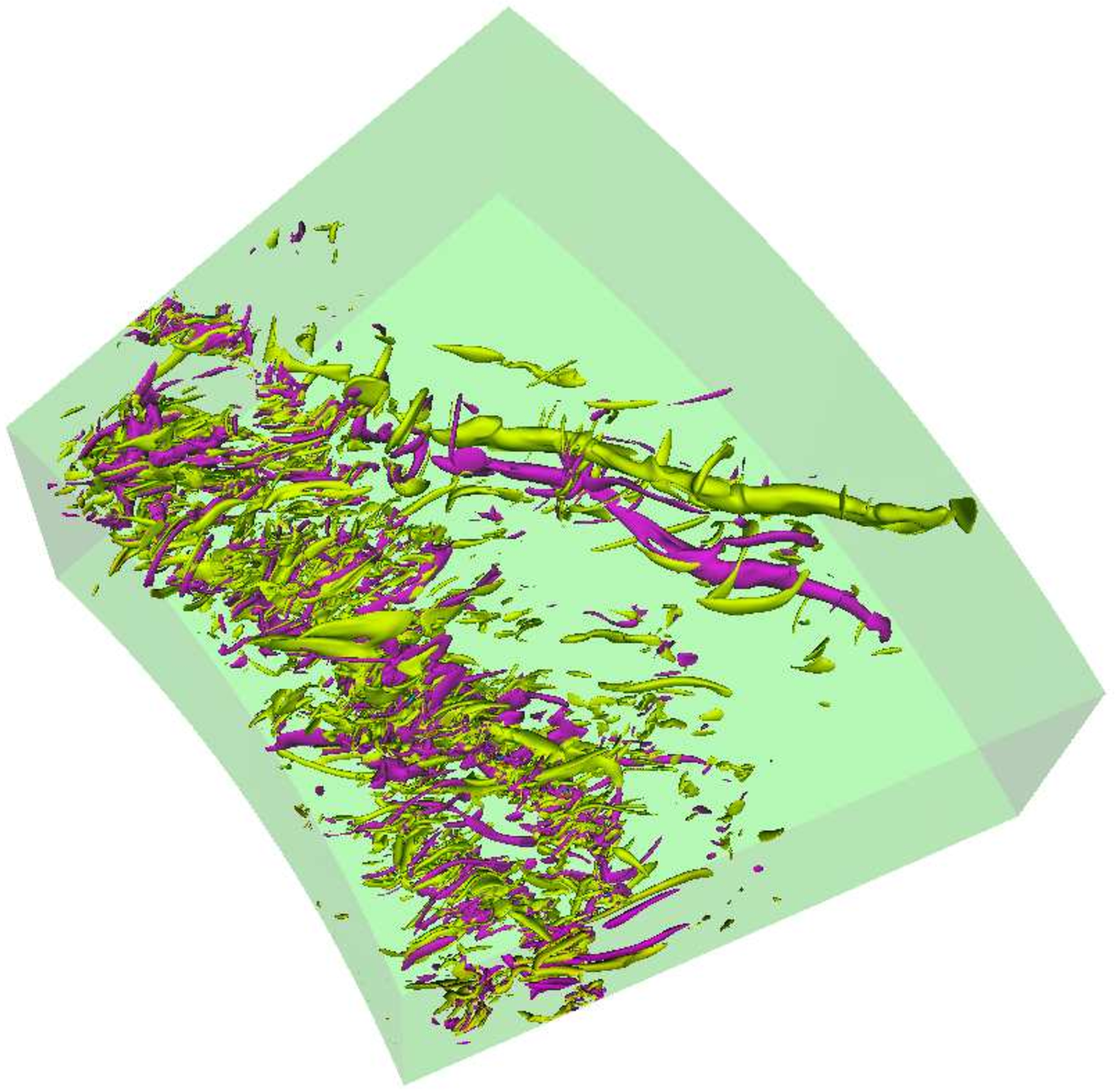}
  \caption{Isosurfaces of the streamwise vorticity in the
    quasi-Keplerian regime at $Re_i=10^5, Re_o=79685$, $\eta=0.71$,
    $\Gamma=0.5$ and $k_\theta=16$. The resolution is $(n_r,n_\theta,n_z)=(1152,
    384, 384)$.}
  \label{fig:highReRun}

\end{figure}

Another test run was performed at $Re_i=10^5, Re_o=79685$ and at
$\eta=0.71$. We used $k_\theta=16$ and axial length $\Gamma=0.5$, with
a spatial resolution $(n_r,n_\theta,n_z)=(1152, 384, 384)$ and
time-step size $\Delta t = 10^{-9}$.  The initial condition at $t=0$
is taken from the optimal initial perturbation, which gives the
maximal transient energy growth~\cite{Maretzke_jfm2014} supplemented
with very small three-dimensional noise. Fig.~\ref{fig:highReRun}
shows the 3D contour plot of the streamwise vorticity,
$\omega_{\theta}=\partial_z u_r-\partial_r u_z$, at $t=5\times10^{-4}$
($\simeq 3.3$ cylinder rotations) which illustrates a transiently
turbulent flow state. The research is still ongoing and will be
disseminated in future publications. We expect that the results will
contribute to clarify the role of pure hydrodynamic turbulence in
astrophysical disks~\cite{Balbus_nature2011}.

\section{Computational efficiency}\label{sec:benchmarks}

\subsection{Benchmark setup}

In this section we report benchmarks results using up to 20\,480
processor cores of an IBM iDataPlex compute cluster with Intel Ivy-Bridge processors 
and a fully nonblocking InfiniBand (FDR~14) fat-tree interconnect. Each shared-memory 
compute node hosts two Intel Xeon E5-2680v2 ten-core processors (CPUs) 
with a clock frequency of 2.8~GHz. 
We employ Intel compilers (version 14.0), and the 
Intel Math Kernel Library (MKL 11.1) for BLAS, LAPACK, and FFT functionality. We have performed two strong scaling studies, which show the scaling of 
the runtime with increasing number of CPU cores for a fixed problem size. 
Two different, representative setups were considered:
\begin{enumerate}[\itshape a\upshape)]
\item
  a ``SMALL'' setup with $(n_r,n_{\theta},n_z)=(32,384,640)$.  This
  setup is used to investigate localized turbulence at the
  transitional stage ($Re\sim \mathcal{O}(10^2)$), where the
  structures inside the turbulence are relatively large.  The
  probability distributions of the splitting time of localized
  turbulent stripe in \S\ref{sec:stripe} are obtained with this
  resolution.
\item
  a ``LARGE'' setup with $(n_r,n_{\theta},n_z)=(2048,384,2048)$.  This
  resolution is representative of our ongoing studies of hydrodynamic
  turbulence in Taylor-Couette flows with quasi-Keplerian velocity
  profiles at Reynolds numbers up to $\mathcal{O}(10^5)$.
\end{enumerate}

For the LARGE setup, a weak scaling study, i.e.\ increasing the problem size
along with the number of CPU cores, is presented in addition.


\subsection{Benchmark results and discussion}

\begin{figure}[!h]
\begin{tabular}{@{\extracolsep{-15pt}}lr}
\includegraphics[width=0.48\textwidth]{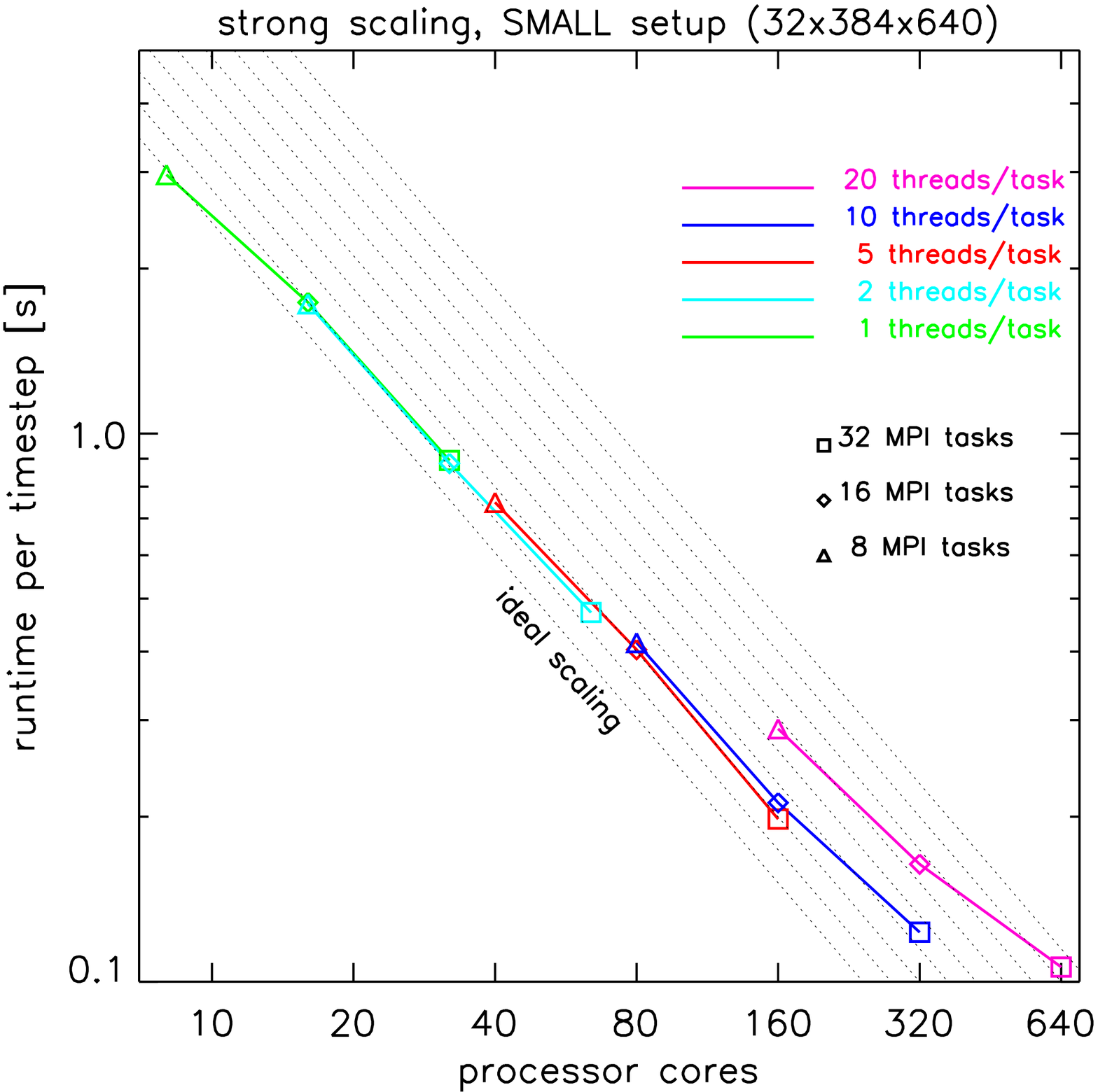} \hfill
\includegraphics[width=0.48\textwidth]{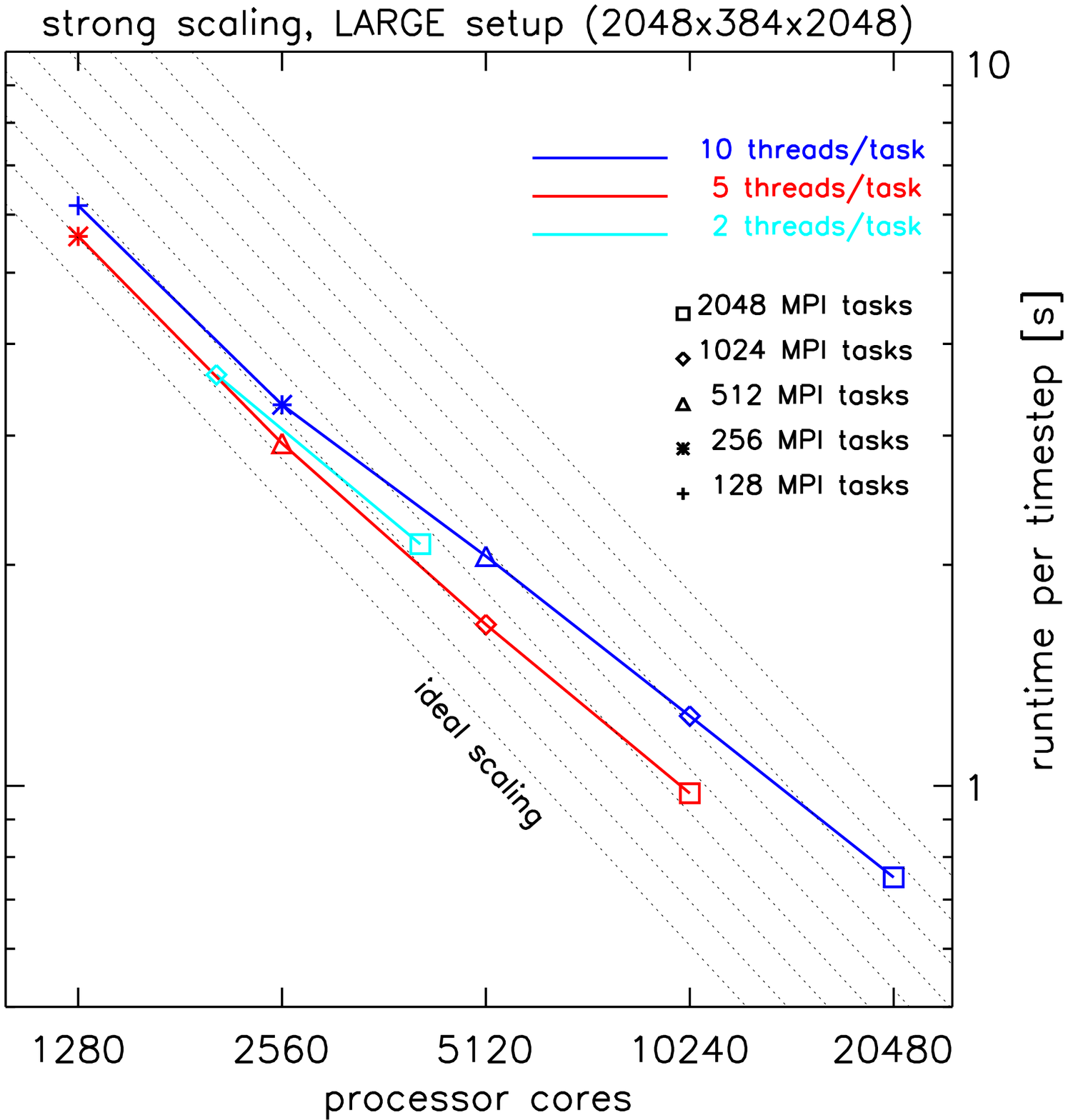}          \\
\includegraphics[width=0.48\textwidth]{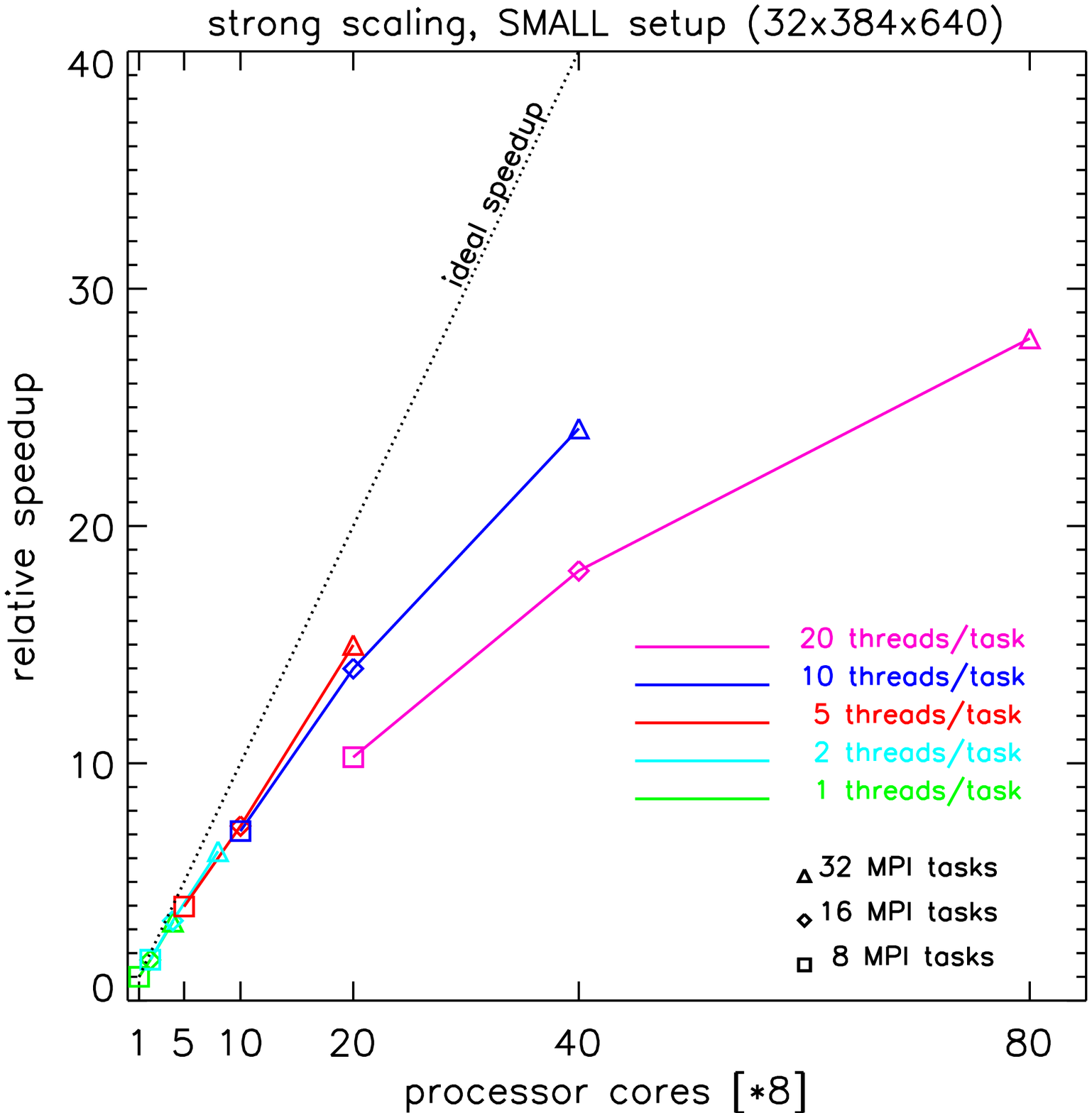} \hfill
\includegraphics[width=0.48\textwidth]{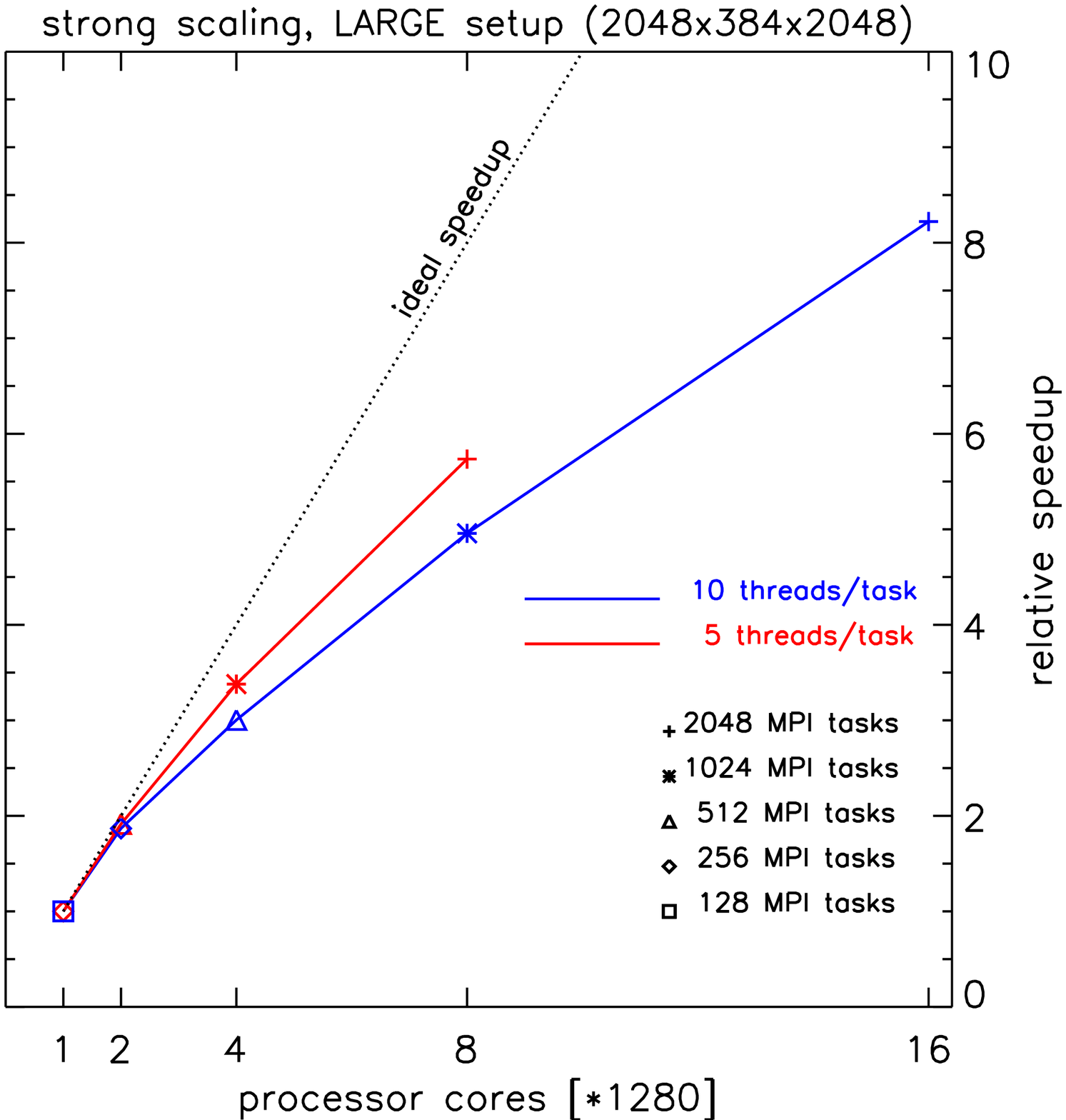}
\end{tabular}
\caption{Runtime per time step (upper row) and relative speedups (bottom row) 
  for the SMALL setup (left panels) and for
  the LARGE setup (right panels) as a function of the number of cores,
  $N_\mathrm{cores}=N_\mathrm{tasks}\cdot N_\mathrm{threads}$. Different
  colors and symbols are used to distinguish runs with different
  numbers of MPI tasks ($N_\mathrm{tasks}$) and OpenMP threads
  ($N_\mathrm{threads}$), respectively. The slope of an ideal scaling
  curve is indicated by dotted lines.}
\label{fig:scaling}
\end{figure}

Fig.~\ref{fig:scaling} provides an overview of the strong 
scalability of the hybrid code. Different colors and symbols are used 
to distinguish runs with different numbers of MPI tasks
($N_\mathrm{tasks}$) and OpenMP threads ($N_\mathrm{threads}$). 
The total number of processor cores is given by $N_\mathrm{cores}=N_\mathrm{tasks}\cdot N_\mathrm{threads}$.

For both setups we achieve scalability up to the maximum
number of cores our parallelization scheme admits on this computing
platform, \textit{i.e.}, \ $N_\mathrm{cores}=32\cdot 20=640$ for the 
SMALL setup, and $N_\mathrm{cores}=2048\cdot 10=20480$ 
for the LARGE setup. The latter number is constrained by the maximum
number of nodes that can be used by a single job.  

  
For the SMALL setup, Fig.~\ref{fig:scaling} (upper left) shows that up
to a number of 10 threads per MPI task the run times for a given number of cores
are virtually the same, independent of the distribution of the resources
to MPI tasks and OpenMP threads. This indicates that the efficiency of our 
coarse-grained OpenMP parallelization is almost the same as the explicit,
MPI-based domain decomposition. 
Moreover, as the results for the LARGE setup 
(Fig.~\ref{fig:scaling}, upper right) show, it can even be more efficient to
use less than the maximum of $n_r$ MPI tasks for a given number of
cores and utilize the resources with OpenMP threads (compare
the cyan and the red symbols at moderate core counts). This is due to the fact that a lower 
number of MPI tasks per node reduces the amount of inter-node
MPI communication (specifically the {\tt MPI\_Alltoall} communication
pattern for the global transpositions). 
Notably, for the LARGE setup, the hybrid code shows nearly
perfect scaling between 1280 and 2560 cores and continues to scale up
to more than 20\,000 processor cores (1024 nodes), albeit with a marginally efficient
speedup of 8 (corresponding to a parallel efficiency of 
slighly more than 50\%) when compared with the baseline run at 1280 cores. A 
floating-point performance of about 20 GFlop/s per compute node is reached which 
is roughly 5\% of the nominal peak performance. The memory footprint
remains below 10 GB per node, making the code well prepared
for future CPU architectures with scarce memory resources \cite{dongarra_hpc2011}.

\bigskip

The details on the absolute run times and the parallel efficiencies of the 
whole code (the bottom row) as well as the individual parts of the algorithm
(cf.\ Section~\ref{sec3:numericalMethod}) are listed in 
Table~\ref{tab:efficiency}. The first column, which
corresponds to a plain MPI-parallelization using the maximum number of
tasks ($N_\mathrm{tasks}=n_r$) for the given setup, is assigned an
efficiency of 100\%, by definition.

\begin{table}[!h]\footnotesize
\begin{center}
\begin{tabular}{|l|lr|lr|lr|lr|lr|}
\hline\hline\multicolumn{11}{|c|}{SMALL setup $(32,384,640)$}\\\hline\hline
\hline
\multicolumn{1}{|r|}{cores ($N_\mathrm{threads}$)} &  
\multicolumn{2}{|c|}{$32 (1)$}& 
\multicolumn{2}{|c|}{$64 (2)$}& 
\multicolumn{2}{|c|}{$160 (5)$}& 
\multicolumn{2}{|c|}{$320 (10)$}& 
\multicolumn{2}{|c|}{$640 (20)$}\\
\hline
 &
$T_1$ [s] & \multicolumn{1}{l|}{$\eta$}& 
$T_2$ [s] & \multicolumn{1}{l|}{$\eta$}& 
$T_5$ [s] & \multicolumn{1}{l|}{$\eta$}& 
$T_{10}$ [s] & \multicolumn{1}{l|}{$\eta$}& 
$T_{20}$ [s] & \multicolumn{1}{l|}{$\eta$}\\
\hline
nonlinear \hfill   (\emph{1})             & 0.696 & 100\% & 0.388 &  90\% & 0.161 &  86\% & 0.101 & 69\% & 0.093 & 37\% \\ 
predictor-corrector \hfill (\emph{2,3,4}) & 0.152 & 100\% & 0.080 &  95\% & 0.033 &  92\% & 0.017 & 89\% & 0.001 & 79\% \\ 
\hline
complete step                             & 0.864 & 100\% & 0.472 &  92\% & 0.200 &  86\% & 0.120 & 72\% & 0.104 & 42\% \\  
\hline
\hline\hline\multicolumn{11}{|c|}{LARGE setup $(2048,384,2048)$}\\\hline\hline
\hline
\multicolumn{1}{|r|}{cores ($N_\mathrm{threads}$)} &  
\multicolumn{2}{|c|}{$2048 (1)$}& 
\multicolumn{2}{|c|}{$4096 (2)$}&      
\multicolumn{2}{|c|}{$10240 (5)$}&
\multicolumn{2}{|c|}{$20480 (10)$}&
\multicolumn{2}{|c|}{}
\\
\hline
 &
$T_1$ [s] & \multicolumn{1}{l|}{$\eta$}& 
$T_2$ [s] & \multicolumn{1}{l|}{$\eta$}& 
$T_5$ [s] & \multicolumn{1}{l|}{$\eta$}& 
$T_{10}$ [s] & \multicolumn{1}{l|}{$\eta$}&
 & \\
\hline
nonlinear      \hfill (\emph{1})          & 3.875 & 100\%& 1.862 & 104\%& 0.870 & 89\%& 0.695 & 70\%&  & \\ 
predictor-corrector \hfill (\emph{2,3,4}) & 0.407 & 100\%& 0.240 &  85\%& 0.097 & 84\%& 0.048 & 85\%&  & \\ 
\hline
complete step                             & 4.325 & 100\%& 2.134 & 101\%& 0.977 & 87\%& 0.751 & 58\%&  & \\  
\hline
\end{tabular}
\caption{Runtime per time step, $T_n$ and parallel efficiency $\eta$ of the
  OpenMP parallelization  
  as a function of the number $N_\mathrm{threads}$ of OpenMP threads per 
  MPI task, using the maximum number of 32 MPI tasks for the SMALL setup, and
  2048 MPI tasks for the LARGE setup, respectively.
  Parallel efficiency is conventionally defined as $\eta:=
  T_1/(n\cdot T_n)$ with $n=N_\mathrm{threads}$.
  Different rows show the contributions of the individual algorithmic 
  steps (numbering in brackets chosen according to Section~\ref{sec3:numericalMethod}) 
  to the total runtime of a complete time step (the bottom row).}\label{tab:efficiency}
\end{center}
\end{table}

For the SMALL setup (the upper part of Table~\ref{tab:efficiency})
we observe good OpenMP efficiency up to 10 threads
(which are pinned to the 10 physical cores of a single CPU socket) per MPI-task for the 
pressure and velocity predictor steps, the corrector step, and also 
the matrix-vector multiplication in the nonlinear part.
When using all 20 cores of a shared-memory node with a single MPI task
(cf.\ the magenta curve in Fig.~\ref{fig:scaling}, left) one notices a
degradation in OpenMP efficiency which is due to memory-bandwidth
limitations, NUMA effects, and limited parallelism in the nonlinear part.
The overall parallel efficiency (the bottom row) can be
considered as very good up to 320 cores,
but gets increasingly bounded by the global transposition ({\tt MPI\_Alltoall}
communication) in the nonlinear part.

\begin{figure}[!ht]
  \centering
  \includegraphics[width=0.8\textwidth]{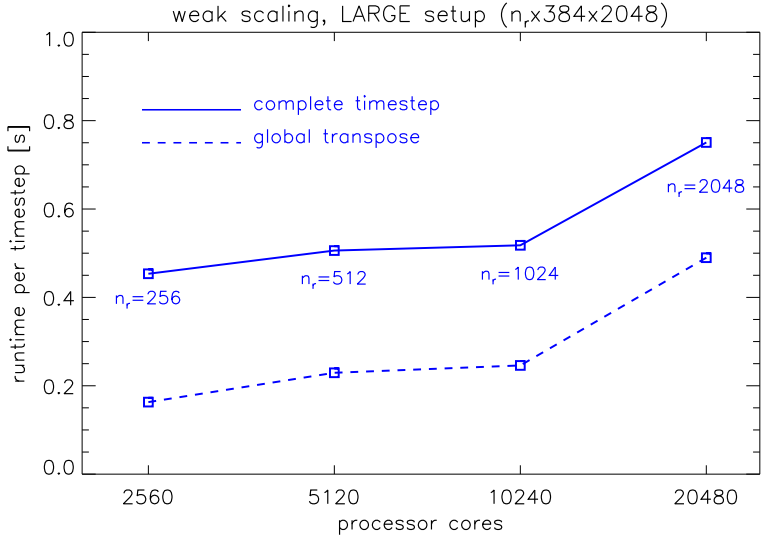}
  \caption{Weak scaling of the total runtime (solid line) and the contribution of the
  global transposition (dashed) for the LARGE setup. The number of cores is related to the number of radial zones as $N_\mathrm{cores}=10\cdot n_r$. In all cases,
  $N_\mathrm{threads}=10$ was used.}
  \label{fig:weakscaling}
\end{figure}

%
For the LARGE setup (the lower part of Table~\ref{tab:efficiency}), 
although the highly scalable linear parts (predictor-corrector, steps 2-4) 
and the matrix-vector multiplications contribute only a minor part to 
the total runtime, the code maintains an excellent OpenMP efficiency up 
to more than 10\,000 cores (87\%). At very high core counts 
the {\tt MPI\_Alltoall} communication dominates the total 
runtime and becomes the major bottleneck for overall scalability.  
This is also apparent in the weak scaling analysis
(cf.\ Fig.~\ref{fig:weakscaling}). The global transposition exhibits
good weak scalability up to 10\,240 cores (512 nodes), and its contribution 
to the total runtime remains subdominant but it seriously impedes the scalability
up to the maximum of 20\,480 cores (1024 nodes).
Note that parts of the global transposition (roughly a third, in terms of
runtime) are performed concurrently with computations and are thus not 
accounted for separately in Fig.~\ref{fig:weakscaling}.

Using an adapted setup of 
$(n_r,n_\theta,n_z)=(1792,384,512)$, we were able to run the code on the largest, fully interconnected partition with 1792 nodes (35\,840 cores) of the 
high-performance computer of the Max-Planck-Society, "Hydra", resulting in a run time of $0.3$~s per time step. Computing times of this order enable us to perform highly resolved simulations 
(\textit{e.g.}\ of Keplerian flows 
which require on the order of a million time steps) within a couple of days.

\bigskip

\section{Conclusion}\label{conclusion}

With the motivation of exploring high-Reynolds-number rotating turbulent flows, we developed a highly-efficient parallel DNS code for Taylor-Couette flows. The incompressible Navier-Stokes equations in cylindrical coordinates are solved in primitive variables by using a projection method proposed by Hugues and Randriamampianina~\cite{Hugues_ijnmf1998}, which is second-order accurate in both pressure and velocity.  This method leads at each time step to the solution of five linear differential equations, either of Poisson or of Helmholtz type, which simplifies significantly the programming of the code. For the spatial discretization, we used a combination of Fourier spectral in axial and azimuthal directions and high-order finite differences in the radial direction, which allow the use of tailored stretched grids. The computing cost scales linearly with the number of grid points in each direction.

In order to reach higher Reynolds numbers and to take full advantage of the modern HPC facilities, the code was parallelized by a hybrid MPI-OpenMP strategy, combining the simplicity of a MPI-based one-dimensional ``slab'' domain decomposition in Fourier space with efficient exploitation of the remaining coarse-grained parallelism by OpenMP threading. Compared to a flat MPI-parallelization, the hybrid code maps more naturally to the current multi-node, multi-core architectures, keeps the number of MPI tasks in the well-manageable regime of a few thousand, and, most importantly, reduces inter-node communications, which improves the overall efficiency and scalability. The strong scaling study which was performed with scientifically relevant setups demonstrates the scalability of the code up to more than 20\,000 processor cores. This allows to perform simulations with much higher resolutions than previously possible. With the current HPC technology, this code pushes the achievable $Re$ to the order of magnitude of $\mathcal{O}(10^5)$ in DNS of Taylor-Couette flow, which therefore opens up the possibility to study quasi-Keplerian flows at experimentally relevant parameters.

The new code was shown to be very accurate in various regimes: laminar Couette flow, wavy vortices, transitional and turbulent flow at high Reynolds number. With the high efficiency of the hybrid parallel scheme, this code possesses great potential to explore the turbulent TCf in a much broader parameter space. 


\section*{Acknowledgments}
We thank Florian Merz (IBM) for optimizing the global transposition
routine. L. Shi and B. Hof acknowledge research funding by Deutsche 
Forschungsgemeinschaft (DFG) under Grant No. SFB963/1 (project A8).
Computations were performed on the HPC system "Hydra" of the Max-Planck-Society
at RZG.





\bibliographystyle{elsarticle-num-names}
\bibliography{./ref_nsCouette.bib}







\end{document}